\documentclass[twocolumn,preprintnumbers,amsmath,prc,amssymb,showpacs]{revtex4}

\usepackage{graphicx}
\usepackage{dcolumn}
\usepackage{bm}
\usepackage{amsmath,amssymb}
\usepackage{lipsum}
\usepackage{slashed}
\usepackage{enumitem}
\usepackage{txfonts}
\usepackage{esint}
\usepackage{xcolor}
\usepackage{mathtools}
\usepackage{transparent}
\usepackage{upgreek}

\begin{document}

\title{Density functional approach to quark matter with confinement and color superconductivity}


\author{Oleksii Ivanytskyi}
\email{oleksii.ivanytskyi@uwr.edu.pl}
\affiliation{Institute of Theoretical Physics, University of Wroclaw, Max Born Pl. 9, 50-204 Wroclaw, Poland}

\author{David Blaschke}
\email{david.blaschke@uwr.edu.pl}
\address{Institute of Theoretical Physics, University of Wroclaw, Max Born Pl. 9, 50-204 Wroclaw, Poland}

\begin{abstract}
We present a novel relativistic density-functional approach to modeling quark matter with a mechanism to mimic confinement. 
The quasiparticle treatment of quarks provides their suppression due to a large quark selfenergy already at the mean-field level. 
We demonstrate that our approach is equivalent to a chiral quark model with medium-dependent couplings. The dynamical restoration of the chiral symmetry is ensured by construction of the density functional. 
Supplemented with the vector repulsion and diquark pairing, the model is applied to construct a hybrid quark-hadron EoS of cold compact-star matter. 
We study the connection of such a hybrid EoS with the stellar mass-radius relation and tidal deformability. The model results are compared to various observational constraints including the NICER radius measurement of PSR J0740+6620 and the tidal deformability constraint from GW170817. 
As a striking result we present selected parametrizations for which the hybrid star sequence reaches masses in the range 2.5 - 2.67 $M_\odot$ of the lighter object in the merger GW190814, which therefore might be interpreted as the heaviest neutron star with to date.
\end{abstract}

\pacs{
97.60.Jd, 
26.60.Kp, 
12.39.Ki 
}
\maketitle

\date{\today}

\section{Introduction} 
\label{sec1}

The quest for a reliable description of the deconfinement transition from hadronic to quark matter is highly actual.
A description from first principles is possible by simulations of quantum chromodynamics (QCD) as a lattice gauge theory, but at present it is restricted to the case of small baryochemical potentials $\mu \lesssim 2.5~T$ \cite{HotQCD:2018pds}, where $T$ is the temperature. 

In order to address the modern challenges of a description of the equation of state of matter created in the beam energy scan programs of present heavy-ion collision (HIC) experiments at CERN and RHIC as well as those planned at FAIR and NICA, and in the interiors of neutron stars as well as their mergers, one needs to develop effective approaches. These should be guided by the lattice QCD studies, neutron star phenomenology and
HIC experiments.
A wide class of such models are the so-called two-phase approaches which model the hadronic and the quark matter phases separately and then perform constructions to describe the transitions between them.

Prominent examples, in particular for hybrid neutron star matter, are the Maxwell construction
for locally conserved charges and the Glendenning construction \cite{Glendenning:1992vb} for several, globally conserved charges. Both are based on the Gibbs principles of phase equilibrium.
They can be unified by invoking the formation of finite-size structures (pasta phases) in the coexistence phase of quark and hadronic matter 
\cite{Heiselberg:1992dx,Yasutake:2014oxa,Maslov:2018ghi}.
The application of these constructions assumes that the EoS models for the two phases are applicable in the region of thermodynamic parameters where the Gibbs conditions for the transition are fulfilled.
Should these conditions not be fulfilled or not make sense, this is a good signature for the inapplicability of the corresponding EoS models in that region. 
In that case, interpolation constructions have recently been suggested \cite{Masuda:2012ed,Alvarez-Castillo:2013spa,Baym:2017whm,Blaschke:2018pva,Abgaryan:2018gqp,Ayriyan:2021prr,Ivanytskyi:2022wln}.
Such an agnostic approach can be understood as a shortcut subsuming three steps in the description of the hadron-to-quark matter transition: 1) the account for the quark substructure of hadrons which leads to quark exchange interactions among hadrons (quark Pauli blocking), 2) the account for confining interactions and color superconductivity and 3) pasta phases in the coexistence region of quark and hadronic matter phases. \\

The present work is devoted to investigating the consequences of the inclusion of diquark interactions and thus the formation of a color superconducting phase on the structure of the EoS and hybrid neutron star properties in a generalization of the density functional approach \cite{Kaltenborn:2017hus} with a confining mean field that was motivated by the string-flip model (SFM) \cite{Horowitz:1985tx,Ropke:1986qs} of quark matter. 
We would like to stress, that strangeness in the density functional approach is an interesting question, which deserves a separate study. The first steps in this direction have been done recently \cite{Blaschke:2022knl}. However, even with the simplifications adopted in that study, the consideration remains quite complicated in the presence of color superconductivity. 
Since for a large class of chiral quark models at zero temperature strange quarks are reported to appear sequentially, after the light ones \cite{Klahn:2006iw,Klahn:2013kga}, we use this as an argument to justify the neglect of strange quarks at the hadron-to-quark matter transition in the present work. This assumption also keeps the considerations sufficiently transparent.

The paper is organized as follows. A general description of the relativistic density functional approach to quark matter within the mean field approximation is given in the next section. Section \ref{sec3} presents a treatment of the pseudoscalar mode needed to 
determine parameters of the model with the known mass and decay constant of the pion that is 
obtained within a Gaussian approximation to the expansion beyond the mean field.
The density functional of the confining interaction potential used in this work is described in Section \ref{sec4}. We discuss normalization of the model in Section \ref{sec5}. In Section \ref{sec6} we give a treatment of cold quark matter.  In Section \ref{sec7} we construct a hybrid quark-hadron EoS and apply it for modeling compact stars with quark matter cores. The conclusions are given in Section \ref{sec8}.

\section{Relativistic density functional approach}
\label{sec2}

A general scheme to treat quark matter within the density functional formalism  under the mean-field approximation has been presented in Ref. \cite{Kaltenborn:2017hus}.
In the case of two quark flavors and three colors the effective interaction in quark matter is accounted for by the potential $\mathcal{U}$ given in terms of quark bilinears of the form $\overline{q}~\hat\Gamma q$. 
Here $q=(u,d)^T$ is a quark spinor and the vertex operator $\hat\Gamma$ acts in color, flavor and Dirac space. 
In the present study we attribute scalar and pseudoscalar interaction channels to the density functional $\mathcal{U}$.
The chiral invariance of the quark interaction is ensured by requiring $\mathcal{U}$ to depend on a chirally symmetric argument. 
In the two flavor case the latter can be chosen as $(\overline{q}q)^2+(\overline{q}i\gamma_5\vec\tau q)^2$. Clearly, $\mathcal{U}$ can include a dependence on other quark bilinears as it was considered in Ref. \cite{Kaltenborn:2017hus}. These complications, however, are not necessary to maintain the proper chiral dynamics of quark matter. Due to this we prefer to consider the simplest $\mathcal{U}$ with the above mentioned chirally symmetric argument. Vector and diquark pairing interaction channels, which are important for the compact star phenomenology, are incorporated to the model within the standard framework \cite{Baym:2017whm}. Thus, the total Lagrangian reads
\begin{eqnarray}
\mathcal{L}&=&\overline{q}(i\slashed\partial-\hat m)q-\mathcal{U}\nonumber\\
\label{I}
&-&G_V(\overline{q}\gamma_\mu q)^2+G_D
(\overline{q}i\gamma_5\tau_2\lambda_A q^c)(\overline{q}^ci\gamma_5\tau_2\lambda_A q),\quad
\end{eqnarray}
where the diagonal matrix $\hat{m}={\rm diag}(m_u,m_d)$ represents current masses of the corresponding states.  Below they are labeled with the subscript index $f=u,d$. 
The charge conjugate quark field is given by $q^c=i\gamma_2\gamma_0\overline{q}^T$, while summation over the color index $A$ in the last term of Eq. (\ref{I}) and elsewhere in this paper is performed for the antisymmetric Gell-Mann matrices $\lambda_A$ with $A=2,5,7$. The strength of the interaction in vector and diquark channels is controlled by the couplings $G_V$ and $G_D$, respectively, 
which are treated as free parameters of the approach.

A Taylor expansion of the interaction potential around the expectation values of the corresponding quark bilinears gives an efficient way of treating the present density functional theory. 
The linear order of the expansion corresponds to the
selfconsistent 
mean-field level \cite{Kaltenborn:2017hus}, which is
applied to all interaction channels 
considered in this work. 
However, 
as an additional constraint to the density functional $\mathcal{U}$ that was not considered in \cite{Kaltenborn:2017hus}, we require the present model to  reproduce the mass and decay constant of the pion in accordance with QCD vacuum phenomenology, thus going beyond the mean-field level in the scalar-pseudoscalar sector.
The corresponding mesonic
correlations can be introduced to the model by expanding $\mathcal{U}$ up to the second order in deviations of quark bilinears from their mean field expectation values $\langle \overline{q}q\rangle$ and $\langle\overline{q}i\gamma_5\vec\tau q\rangle=0$. Below all the quantities defined at these expectation values are denoted with the subscript index $``MF"$. Then, the second order expanded potential energy density functional
obtains the form
\begin{eqnarray}
\mathcal{U}^{(2)}&=&\mathcal{U}_{MF}+
\left(\overline{q}q-\langle\overline{q}q\rangle\right)\Sigma_{MF}\nonumber\\
\label{II}
&-&G_S(\overline{q}q-\langle\overline{q}q\rangle)^2-
G_{PS}(\overline{q}i\gamma_5\vec\tau q)^2,
\end{eqnarray}
where the expansion coefficients are
\begin{eqnarray}
\label{III}
\Sigma_{MF}&=&\frac{\partial\mathcal{U}_{MF}}{\partial\langle\overline{q}q\rangle},\\
\label{IV}
G_S&=&-\frac{1}{2}
\frac{\partial^2\mathcal{U}_{MF}}{\partial\langle\overline{q}q\rangle^2},\\
\label{V}
G_{PS}&=&-\frac{1}{6}
\frac{\partial^2\mathcal{U}_{MF}}{\partial\langle\overline{q}i\gamma_5\vec\tau q\rangle^2}.
\end{eqnarray}
They represent the mean-field part of the quark self-energy $\Sigma_{MF}$, while $G_S$ and $G_{PS}$ are medium dependent effective couplings in the scalar and pseudoscalar channels, respectively. 
The mean field self-energy of quarks renormalizes their mass resulting in the medium-dependent effective mass $m^*_f=m_f+\Sigma_{MF}$. It is important to note that the scalar $G_S$ and pseudoscalar $G_{PS}$ couplings coincide up to constant positive factors with the second derivatives of the Lagrangian $\mathcal{L}$ with respect to scalar $\langle\overline{q}q\rangle$ and pseudoscalar $\langle\overline{q}i\gamma_5\vec\tau q\rangle$ condensates. Positiveness of these derivatives provides that mean-field solution corresponds to a minimum of the thermodynamic potential. Thus, we require $G_S>0$ and $G_{PS}>0$.

The Lagrangian (\ref{I}) with the density functional expanded up to the second order terms (\ref{II}) yields $\mathcal{L}^{(2)}=\mathcal{L}_{\mathcal{U}=0}-\mathcal{U}^{(2)}$, 
which is quadratic in all quark bilinears present in the model. 
With this Lagrangian the partition function of the present model can be given the form of the functional integral over the quark fields
\begin{eqnarray}
\label{VI}
\mathcal{Z}=\int\mathcal{D}\overline{q}~\mathcal{D}q\exp\left[\int dx_E~
\left(\mathcal{L}^{(2)}+q^+\hat\mu q\right)
\right].
\end{eqnarray}
Hereafter integration over the Euclidean space-time $dx_E=d\tau~d\bf x$ is limited to the system volume $V=\int d\bf x$ and inverse temperature $\frac{1}{T}\equiv\beta=\int_0^\beta d\tau$. 
The diagonal matrix $\hat\mu={\rm diag}(\mu_u,\mu_d)$ acting in the flavor space represents the quark chemical potentials $\mu_f=\frac{\mu_B}{3}+Q_f\mu_Q$. They are expressed in terms of the baryonic 
($\mu_B$) and electric ($\mu_Q$) chemical potentials as well as the baryonic (1/3) and electric ($Q_f$) charges of the quarks.
The partition function provides direct access to the thermodynamic potential $\Omega=-(T/V)\ln\mathcal{Z}$ from which all equations of state can be derived.

To proceed further we bosonize $\mathcal{L}^{(2)}$ by means of the Hubbard-Stratonovich transformation in the spirit of Ref. \cite{Blaschke:2013zaa}. 
This introduces collective scalar $\sigma$, pseudoscalar $\vec\pi$, vector $\omega_\mu$ and complex scalar diquark $\Delta_A$ fields, coupled to $\overline{q}q-\langle\overline{q}q\rangle$, $\overline{q}i\gamma_5\vec\tau q$, $\overline{q}\gamma_\mu q$ and $\overline{q}i\gamma_5\tau_2\lambda_A q$, respectively. Then the partition function becomes 
\begin{eqnarray}
\label{VII}
\mathcal{Z}&=&\int\mathcal{D}\sigma~\mathcal{D}\vec\pi~\mathcal{D}\omega_\mu~\mathcal{D}\Delta_A~\mathcal{D}\Delta_A^*~\mathcal{D}\overline{q}~\mathcal{D}q\nonumber\\
&\times&\exp\left[\int dx_E\left(\mathcal{L}^{bos}+q^+\hat\mu q\right)\right].
\end{eqnarray}
The bosonized Lagrangian $\mathcal{L}^{bos}$ can be expressed through the Nambu-Gorkov bispinor
quark fields
$\mathcal{Q}^T=\frac{1}{\sqrt{2}}(q~q^c)$ as
\begin{eqnarray}
\mathcal{L}^{bos}+q^+\hat\mu q&=&\overline{\mathcal{Q}}\mathcal{S}^{-1}\mathcal{Q}-\mathcal{U}_{MF}
+\langle\overline{q}q\rangle\Sigma_{MF}+
\sigma\langle\overline{q}q\rangle\nonumber\\
\label{VIII}
&-&\frac{\sigma^2}{4G_S}-\frac{\vec\pi^2}{4G_{PS}}+
\frac{\omega_\mu\omega^\mu}{4G_V}-\frac{\Delta_A^*\Delta_A^{}}{4G_D}.
\end{eqnarray}
Here the inverse Nambu-Gorkov quark propagator is
\begin{eqnarray}
\label{IX}
\mathcal{S}^{-1}=\left(
\begin{array}{l}
S^{-1}_+-\sigma-i\gamma_5\vec\tau \cdot\vec\pi\hspace*{1cm}
i\Delta_A^{}\gamma_5\tau_2\lambda_A\\\hspace*{.5cm}
i\Delta_A^*\gamma_5\tau_2\lambda_A\hspace*{1cm}S^{-1}_--\sigma-i\gamma_5\vec\tau^T\cdot\vec\pi
\end{array}\right),
\end{eqnarray}
with $S^{-1}_\pm=i\slashed\partial\pm\slashed\omega-m^*\pm\gamma_0\hat\mu$. 
The quark fields in Eq.~(\ref{VII}) can be integrated out since $\mathcal{L}^{bos}$ is quadratic with respect to them. 
This yields the partition function
\begin{eqnarray}
\label{X}
\mathcal{Z}&=&\int\mathcal{D}\sigma~\mathcal{D}\vec\pi~\mathcal{D}\omega_\mu~\mathcal{D}\Delta_A~\mathcal{D}\Delta_A^*
\nonumber\\
&&\exp\left[\frac{1}{2}{\rm Tr}\ln(\beta\mathcal{S}^{-1})
-\int dx_E\Biggl(\mathcal{U}_{MF}
-\langle\overline{q}q\rangle\Sigma_{MF}
\right.\nonumber\\
&&\left.-\sigma\langle\overline{q}q\rangle+
\frac{\sigma^2}{4G_S}+\frac{\vec\pi^2}{4G_{PS}}-
\frac{\omega_\mu\omega^\mu}{4G_V}+\frac{\Delta_A^*\Delta_A^{}}{4G_D}\Biggl)\right].
\end{eqnarray}
From now on we prefer to work in the momentum representation rather than in the coordinate one. 
Thus, for the quark propagator in Eq. (\ref{IX}) we use the Fourier transformed bosonic fields and $S^{-1}_\pm=\slashed k-\hat m^*$ with $k_0=iz_n\pm\hat\mu^*$ and $z_n=(2n+1)\pi T$ being a fermionic Matsubara frequency. 
The trace in Eq.~(\ref{X}) is performed in the Nambu-Gorkov, Dirac, flavor, and color spaces, as well as over three-momentum and Matsubara indices. We note that the factor $\frac{1}{2}$ compensates the artificial doubling of quark degrees of freedom in the Nambu-Gorkov formalism. 

This work is focused on the mean field treatment of the present model. 
In this case the functional integration over the bosonic fields is dropped, while the fields themselves are replaced by their values at the extremum of the Euclidean action. 
These expectation values can be deduced from the gap equations obtained by averaging the corresponding Euler-Lagrange equations,
\begin{eqnarray}
\label{XI}
\sigma&=&-2G_{S}(\overline{q}q-\langle\overline{q}q\rangle),\\
\label{XII}
\vec\pi&=&-2G_{PS}\overline{q}i\gamma_5\vec\tau q,\\
\label{XIII}
\omega_\mu&=&-2G_V\overline{q}\gamma_\mu q,\\
\label{XIV}
\Delta_A&=&2G_D\overline{q}^ci\tau_2\gamma_5\lambda_A q.
\end{eqnarray}
We note that the equation for $\Delta_A^*$ is equivalent to the hermitean conjugate of Eq.~(\ref{XIV}). It follows from the equations for (pseudo)scalar fields that $\langle\sigma\rangle=\langle\vec\pi\rangle=0$ under the mean field approximation. 
Therefore, the contribution of $\sigma$ and $\vec\pi$ to the thermodynamic properties of the present model is neglected at this level. 
Below, however, we study the pseudoscalar mode within the Gaussian approximation in order to normalize our approach to the vacuum phenomenology of QCD. 
The expectation values of the vector and diquark fields can be obtained by averaging Eqs. (\ref{XIII}) and (\ref{XIV}), respectively. 
By a proper Lorentz transform, the vector field attains the form $\langle\omega_\mu\rangle=g_{\mu0}\omega$. 
The constant $\omega=-2G_V\langle q^+ q\rangle$ is absorbed to the effective chemical potential of quarks $\mu^*_f=\mu_f+\omega$. 
Furthermore, we perform a global color rotation leaving $\Delta_2^{}$ the only diquark field with non vanishing expectation value. 
Only its modulus $\Delta=2G_D|\langle\overline{q}^ci\tau_2\gamma_5\lambda_2 q\rangle|$ appears in the expressions for thermodynamic quantities.
This non-vanishing diquark field is conjugated to Gell-Mann matrix $\lambda_2$ pairing only red and green quark states, while leaving the blue one unpaired. We label these states with the subscript index $c=(r,g,b)$. The corresponding single particle energies shifted by the effective chemical potential $\mu_f^*$ can be found from the condition $\det(\mathcal{S}^{-1}_{MF})=0$ as
\begin{eqnarray}
\label{XV}
\epsilon^\pm_{{\bf k}fc}={\rm sgn}(\epsilon_{{\bf k}f}\mp\mu_f^*)\sqrt{(\epsilon_{{\bf k}f}\mp\mu_f^*)^2+\Delta^2_c},
\end{eqnarray}
where the subscript index $\bf k$ labels quark momentum states, $\epsilon_{{\bf k}f}=\sqrt{{\bf k}^2+m_f^{*2}}$ and $\Delta_r=\Delta_g=\Delta$, $\Delta_b=0$ are introduced for unifying the notations. 
Hereafter, the superscript indices $``+"$ and $``-"$ correspond to quarks and antiquarks, respectively. 
It follows from this expression that finite $\Delta$ generates a gap in the dispersion relation of red and green quarks at $\mu_f^*>m_f^*$ and antiquarks of the same colors at $\mu_f^*<-m_f^*$.

With the above notations, the mean-field thermodynamic potential can be written as 
\begin{eqnarray}
\label{XVI}
\Omega=\Omega_q+\mathcal{U}_{MF}-\langle\overline{q}q\rangle\Sigma_{MF}-\frac{\omega^2}{4G_V}+\frac{\Delta^2}{4G_D}.
\end{eqnarray}
Its quark part
\begin{eqnarray}
\Omega_q&=&-\frac{T}{2 V}
{\rm Tr}\ln(\beta\mathcal{S}^{-1}_{MF})\nonumber\\
\label{XVII}
&=&-2\sum_{f,c,a=\pm}\int\frac{d{\bf k}}{(2\pi)^3}
\left[\frac{g_{\bf k}}{2}\epsilon_{{\bf k}fc}^a
-T\ln\left(1-f^a_{{\bf k}fc}\right)\right]
\end{eqnarray}
includes the single particle distribution function
$f_{{\bf k}fc}^{\pm}=\left[e^{\beta\epsilon_{{\bf k}fc}^\pm}+1\right]^{-1}$.
The zero point terms in Eq. (\ref{XVII}) are regularized by the formfactor  
\begin{eqnarray}
\label{XVIII}
g_{\bf k}=\exp\left[-\left(\frac{\bf k^2}{\Lambda^2}\right)^\xi\right]
\end{eqnarray} 
with constant $\xi>{1}/{2}$ and $\Lambda$ defining a momentum scale. This formfactor has an inflection point at $|{\bf k}|=\Lambda\left(1-\frac{1}{2\xi}\right)^{1/{2\xi}}$. In this work we use its Gaussian form with $\xi=1$.
Regularization with a sharp momentum cutoff corresponds to $\xi\rightarrow\infty$ yielding $g_{\bf k}\rightarrow
\theta(\Lambda-|{\bf k}|)$. 
However, we omit this sharp regularization scheme since it leads to unphysical irregularities in the behavior of various thermodynamic quantities of the 2SC phase. In order to show this we consider the number density $\langle f^+ f\rangle=-{\partial\Omega_q}/{\partial\mu_f}$ for a given quark flavor $f$.
Differentiation of ${\rm sgn}(\epsilon_{{\bf k}f}\mp\mu_f^*)$ in the dispersion relation of (anti)quarks produces the Dirac delta-function $\delta(\epsilon_{{\bf k}f}\mp\mu_f^*)$. The part of $\langle f^+ f\rangle$, which include this delta-function, is
\begin{eqnarray}
\label{XIX}
\delta\langle f^+f\rangle&=&2\sum_{c,a=\pm}a\int\frac{d{\bf k}}{(2\pi)^3}
\left(f_{{\bf k}fc}^a-\frac{g_{\bf k}}{2}\right)
2\Delta_c\delta(\epsilon_{{\bf k}fb}^a)\nonumber\\
&=&\frac{2\mu_f^* k_f\Delta}{\pi^2}(1-g_{\bf k})_{|{\bf k}|=k_f},
\end{eqnarray}
where $k_f=\theta\left(\mu_f^{*2}-m_f^{*2}\right)\sqrt{\mu_f^{*2}-m_f^{*2}}$ is the Fermi momentum of a given quark flavor. 
From this expression it is clear that any discontinuity of the formfactor $g_{\bf k}$ translates to behavior of quark number density if $\Delta\neq0$. 
The same conclusion holds for any thermodynamic quantity of the 2SC phase. 
This explains the choice of the momentum formfactor $g_{\bf k}$.

The mean-field equation for the zeroth component of vector field is
\begin{eqnarray}
\omega&=&2G_V\sum_f\frac{\partial\Omega_q}{\partial\mu_f}\nonumber\\
\label{XX}
&=&4G_V\hspace*{-.2cm}\sum_{f,c,a=\pm}\hspace*{-.1cm}a\int\frac{d{\bf k}}{(2\pi)^3}
\left(\frac{g_{\bf k}}{2}-f_{{\bf k}fc}^a\right)
\left(2\Delta_c\delta(\epsilon_{{\bf k}fb}^a)+\frac{\epsilon_{{\bf k}fb}^a}{\epsilon_{{\bf k}fc}^a}\right).
\nonumber\\
\end{eqnarray}
The pairing gap equation can be derived in a similar way
\begin{eqnarray}
\label{XXI}
\Delta&=&-2G_D\sum_f
\frac{\partial\Omega_q}{\partial\Delta}
\nonumber\\
&=&4G_D\sum_{f,c,a=\pm}\int\frac{d{\bf k}}{(2\pi)^3}
\left(\frac{g_{\bf k}}{2}-f_{{\bf k}fc}^a\right)
\frac{\Delta_c}{\epsilon_{{\bf k}fc}^a}.
\end{eqnarray}
This equation has two solutions. 
The trivial one, $\Delta=0$, corresponds to the normal phase. Under suitable conditions, Eq. (\ref{XXI}) also has a nontrivial solution $\Delta\neq 0$, signaling the formation of a color superconducting phase of quark matter. Due to the presence of an attractive quark-quark interaction, according to the Cooper theorem
\cite{Cooper:1956zz}, the occurrence of such a phase is inevitable at sufficiently low temperatures and high quark number densities.
Finally, the equation for the chiral condensate reads
\begin{eqnarray}
\langle\overline{q}q\rangle&=&
\sum_f\frac{\partial\Omega_q}{\partial m_f}\nonumber\\
\label{XXII}
&=&2\hspace*{-.1cm}\sum_{f,c,a=\pm}\int\frac{d{\bf k}}{(2\pi)^3}
\left(f_{{\bf k}fc}^a-\frac{g_{\bf k}}{2}\right)
\left(2\Delta_c\delta(\epsilon_{{\bf k}fb}^a)+\frac{\epsilon_{{\bf k}fb}^a}{\epsilon_{{\bf k}fc}^a}\right)\frac{m^*}{\epsilon_{{\bf k}f}}.
\nonumber\\
\end{eqnarray}

Eqs. (\ref{XX})-(\ref{XXII}) provide the stationarity of the thermodynamic potential. 
Having these conditions solved, the pressure, the number density of a given quark flavor, the entropy density and the energy density can be found using standard thermodynamic identities such as 
$p=-\Omega+\Omega_0$, 
$n_f={\partial p}/{\partial\mu_f}$, $s={\partial p}/{\partial T}$, $\varepsilon=\sum_f\mu_f n_f+Ts-p$, respectively. Hereafter, the subscript $``0"$ denotes the quantities defined in the vacuum, i.e. at $\hat\mu=T=0$.

\section{Scalar and pseudoscalar modes}
\label{sec3}

We now proceed to the analysis of the (pseudo)scalar fluctuation modes
$\Sigma=\Sigma_\sigma+\Sigma_\pi$ with
\begin{eqnarray}
\label{XXIII}
\Sigma_\pi=-\left(
\begin{array}{l}
i\gamma_5\vec\tau\vec\pi\hspace*{.6cm}0\\
\hspace*{.3cm}0\hspace*{.5cm}i\gamma_5\vec\tau^T\vec\pi
\end{array}\right),\quad
\Sigma_\sigma=-\left(
\begin{array}{l}
\sigma\hspace*{.2cm}0\\
0\hspace*{.2cm}\sigma\end{array}\right),
\end{eqnarray}
that appear beyond the mean field approximation 
as dressing of the inverse Nambu-Gorkov quark propagator
$\mathcal{S}^{-1}=\mathcal{S}^{-1}_{MF}+\Sigma$. 
Within the Gaussian approximation, $\Sigma$ is assumed to be small compared to $\mathcal{S}^{-1}_{MF}$. Thus, the logarithm of the inverse dressed quark propagator 
can be expanded as
\begin{eqnarray}
\label{XXIV}
\ln\left(\beta\mathcal{S}^{-1}\right)&=&
\ln\left(\beta\mathcal{S}^{-1}_{MF}\right)+
\ln\left(1+\mathcal{S}_{MF}\Sigma\right)\nonumber\\
&=&\ln\left(\beta\mathcal{S}^{-1}_{MF}\right)+\mathcal{S}_{MF}\Sigma-\frac{\left(\mathcal{S}_{MF}\Sigma\right)^2}{2}+\mathcal{O}\left(\Sigma^3\right).\quad
\end{eqnarray}
The trace of the pion contribution in the first order term in this expansion vanishes due to the 
tracelessness of the flavor Pauli matrices. 
The trace of $\mathcal{S}_{MF}\Sigma_\sigma$ exactly cancels the term $\int dx_E~\sigma\langle\overline{q}q\rangle$ in Eq. (\ref{X}) and, therefore, both terms vanish. 
The pion-sigma cross contribution $\propto\Sigma_\sigma\Sigma_\pi$ in the second order term also vanishes due to traceless character of the Pauli matrices. 
The remaining second order terms $\sim\Sigma_\sigma^2$ and $\sim\Sigma_\pi^2$ are quadratic in the quark propagator and represent one-loop polarization operators of scalar and pseudoscalar modes at four-momentum $p$,
\begin{eqnarray}
\label{XXV}
\Pi_\pi&=&-\frac{1}{2\beta V}{\rm Tr}(i\gamma_5\mathcal{S}_{MF})^2=-\frac{\langle\overline{q}q\rangle}{m^*}+p^2\mathcal{I}(p^2),\\
\label{XXVI}
\Pi_\sigma&=&-\frac{1}{2\beta V}{\rm Tr}(\mathcal{S}_{MF})^2=-\frac{\langle\overline{q}q\rangle}{m^*}+
(p^2-4m^{*2})\mathcal{I}(p^2).\quad
\end{eqnarray}
Here the integral $\mathcal{I}(p^2)$ has been introduced as
\begin{eqnarray}
\label{XXVII}
\mathcal{I}(p^2)=2\sum_{f,c,a=\pm}\int\frac{d{\bf k}}{(2\pi)^3}\frac{g_{\bf k}-2f_{{\bf k}fc}^a}{2\epsilon_{_{\bf k}f}(4\epsilon_{{\bf k}f}^2-p^2)}.
\end{eqnarray}
For the sake of simplicity we evaluated $\Pi_\sigma$ and $\Pi_\pi$ at $\Delta=0$, which is sufficient to define the vacuum properties of the (pseudo)scalar modes.
The details of the derivation can be found in Refs.
\cite{Klevansky:1992qe,Blaschke:2013zaa,Hufner:1994ma,Zhuang:1994dw,Buballa:2003qv}.
These polarization operators enter the inverse of the Fourier transformed mesonic propagators as
\begin{eqnarray}
\label{XXVIII}
D_\sigma^{-1}=\frac{1}{2G_{S}}-\Pi_\sigma,\quad
D_\pi^{-1}=\frac{1}{2G_{PS}}-\Pi_\pi.
\end{eqnarray}
These two-dimensional matrices act in the momentum-Matsubara space. 
The poles of $D_\sigma$ and $D_\pi$ at ${\bf p}=0$ define mesonic masses being solutions of
\begin{eqnarray}
\label{XXIX}
M_\sigma^2&=&\left(\frac{1}{2G_{S}}+\frac{\langle\overline{q}q\rangle}{m^*}\right)\mathcal{I}^{-1}(M_\sigma^2)+4m^{*2},\\
\label{XXX}
M_\pi^2&=&\left(\frac{1}{2G_{PS}}+\frac{\langle\overline{q}q\rangle}{m^*}\right)\mathcal{I}^{-1}(M_\pi^2).
\end{eqnarray}
With the effective quark mass expressed as $m^*=m-2G_{PS}\langle\overline{q}q\rangle$,
the equation for $M_\pi^2$ can be given a form of the Gell-Mann-Oakes-Renner relation \cite{Gell-Mann:1968hlm}
$M_\pi^2F_\pi^2=-m\langle\overline{q}q\rangle$
with the pion decay constant
\begin{eqnarray}
\label{XXXI}
F_\pi^2=m^*(m^*-m)\mathcal{I}(M_\pi^2).
\end{eqnarray}
Eqs. (\ref{XXIX}) - (\ref{XXXI}) give a direct access to the vacuum phenomenology of QCD.

\section{Density functional}
\label{sec4}

\begin{figure}[t]
\includegraphics[width=0.9\columnwidth]{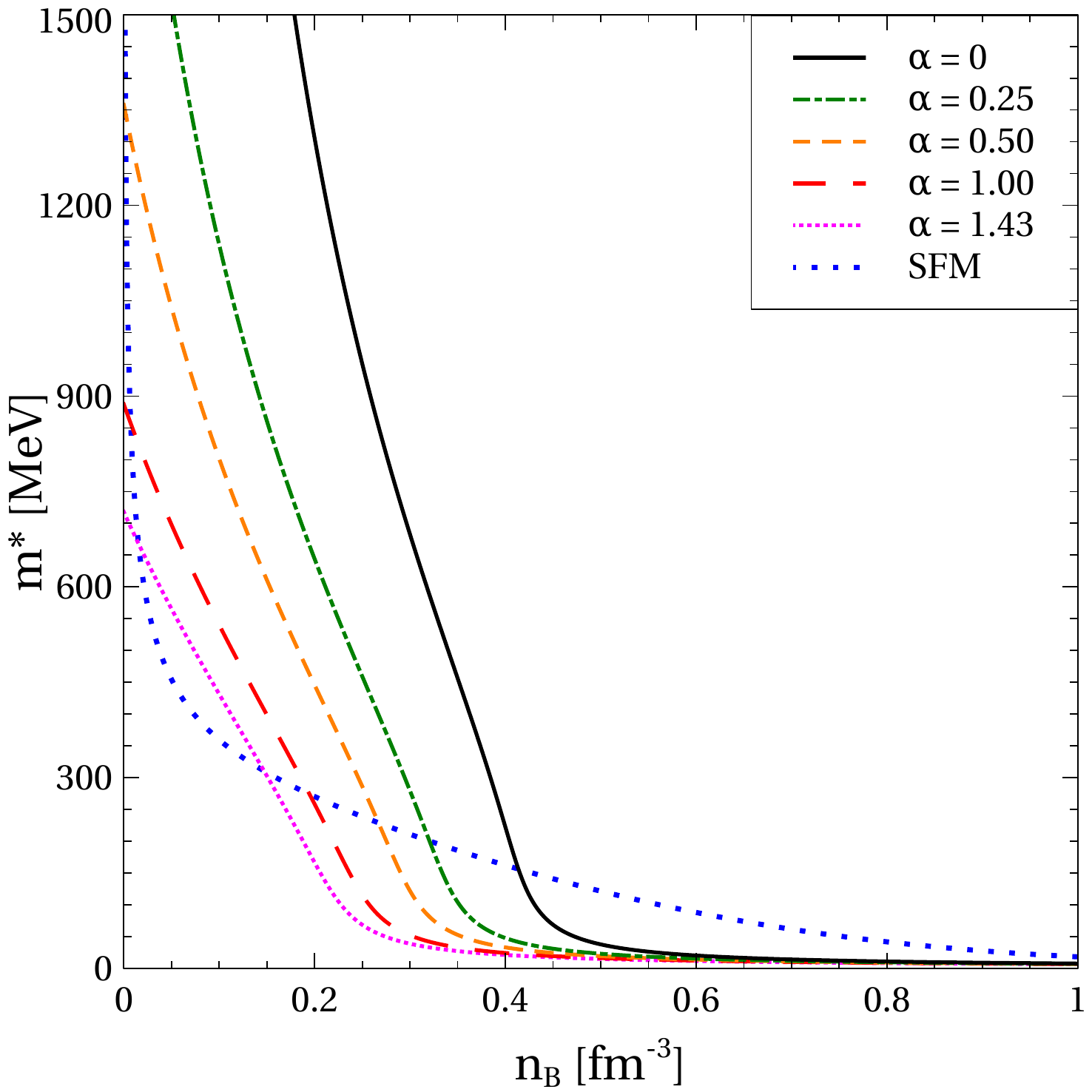}
\caption{Effective quark mass $m^*$ as functions of baruonic density $n_B$ at different values of the parameter $\alpha$. Calculations are performed for cold symmetric quark matter and $G_V=G_D=0$ while the rest of the model parameters except $\alpha$ have the values from Table \ref{table1}. Blue long dotted line is obtained within the SFM with $\alpha_{SFM}=0.39~fm^{-3}$ (see Ref. \cite{Kaltenborn:2017hus} for details).}
\label{fig1new}
\end{figure}

The confining aspect of the quark interaction is modeled within our approach by a large and positive scalar quark self-energy $\Sigma_{MF}$ in the confining region. 
The SFM argument implies that the quark mass grows proportional to the mean interquark separation or, equivalently, $\Sigma\propto D_0 (q^+q)^{-{1}/{3}}$
\cite{Horowitz:1985tx,Ropke:1986qs},
with $D_0$ being proportional to the string tension.
It is worth mentioning, that an alternative prescription for the low density behavior of the effective quark mass implies $m^*\propto\langle q^+q\rangle^{-1}$ \cite{Fowler:1981rp}. For the reader's convenience, we demonstrate in the Appendix A how this result can be obtained under some simplifications when compared to our approach. According to Eq. (\ref{III}), the
SFM behavior of $\Sigma$ corresponds to $\mathcal{U}\propto(q^+q)^{\frac{2}{3}}$. 
In the confining region of low temperature and chemical potential, the quark number density and the chiral condensate are connected by the approximate relation $\langle\overline{q}q\rangle\simeq\langle\overline{q}q\rangle_0+\langle q^+q\rangle$. 
This motivates 
$\mathcal{U}\propto(\langle\overline{q}q\rangle_0-\overline{q}q)^{\frac{2}{3}}$. 
At the same time, 
the chiral invariance of the Lagrangian (\ref{I}) 
is ensured within our approach 
by the appropriate choice of the argument 
$(\overline{q}q)^2+(\overline{q}i\gamma_5\vec\tau q)^2$
in the ansatz for the density functional $\mathcal{U}$.
In order to provide such property of the chirally symmetric interaction mimicking confinement by a significant growth of the quark self-energy we consider the parameterization
\begin{eqnarray}
\label{XXXII}
\mathcal{U}=D_0\left[(1+\alpha)\langle \overline{q}q\rangle_0^2
-(\overline{q}q)^2-(\overline{q}i\gamma_5\vec\tau q)^2\right]^\varkappa,
\end{eqnarray}
where $D_0$ is the coupling strength and $\alpha$ is a constant parameter regulating the value of the quark mass in the vacuum.
Setting $\varkappa={1}/{3}$ is in accordance to the SFM prescription adopted in \cite{Kaltenborn:2017hus}. At zero baryonic density, Eq. (\ref{XXXII}) reproduces the parameterization used in Ref. \cite{Kaltenborn:2017hus} if the no-sea approximation is applied which neglects the quark vacuum term in the thermodynamic potential and the pseudoscalar interaction channel is neglected. 
Indeed, in this approximation chiral condensate coincides up to the sign with the quark scalar density, i.e.  $n_s=-\langle \overline{q}q\rangle$ and $\langle \overline{q}q\rangle_0=0$. This makes Eq. (\ref{XXXII}) of the present work equivalent to Eq. (23) of Ref. \cite{Kaltenborn:2017hus}. The role of $\alpha$ becomes clear from considering effective quark mass in the vacuum when $\langle \overline{q}q\rangle=\langle \overline{q}q\rangle_0$ and $\langle\overline{q}i\gamma_5\vec\tau q\rangle=0$. With the definition $\hat m^*=\hat m+\Sigma_{MF}$ and the mean field self energy of quarks given by Eq. (\ref{III}) we obtain
\begin{eqnarray}
\label{XXXIII}
\hat m^*_0=\hat m+\frac{2}{3}D_0
\alpha^{-2/3}\langle\overline{q}q\rangle_0^{-1/3},
\end{eqnarray}
At high $\alpha$ effective quark mass coincides with the current one, which means that quark interaction in the (pseudo)scalar channels is suppressed. On the other hand, at $\alpha=0$ vacuum effective mass diverges making any excitations of quark degrees of freedom energetically disfavored. 
At small positive values of this parameter $\hat m^*_0$ remains finite but large leading to sufficient suppression of quarks. 
This simple physical mechanism provides an effective phenomenological confinement within the present model. 
In order to illustrate the role of $\alpha$ we show in Fig. \ref{fig1new} the dependence of the effective quark mass on the baryon density found for cold symmetric quark matter in the absence of diquark pairing.
This $m^*$ decreases with growing $n_B$ and vanishes asymptotically. Larger $\alpha$ parameters correspond to a faster decrease of $m^*$ and to its smaller vacuum value, which remains finite at all $\alpha>0$. 
At $\alpha=0$, the effective quark mass in the vacuum diverges similarly to the case of the SFM \cite{Kaltenborn:2017hus} shown for the sake of comparison.

\begin{figure}[thb]
\includegraphics[width=0.9\columnwidth]{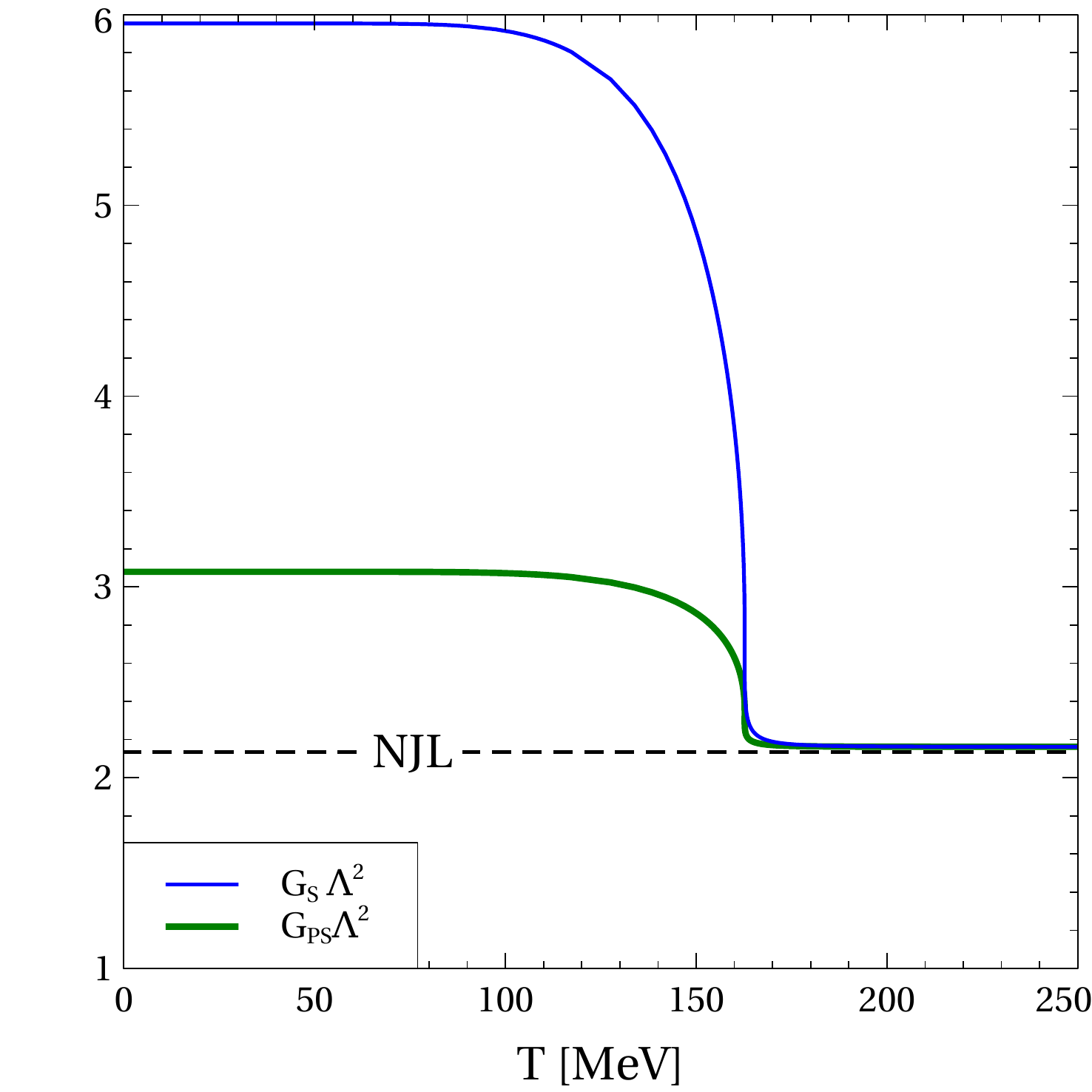}\\
\includegraphics[width=0.9\columnwidth]{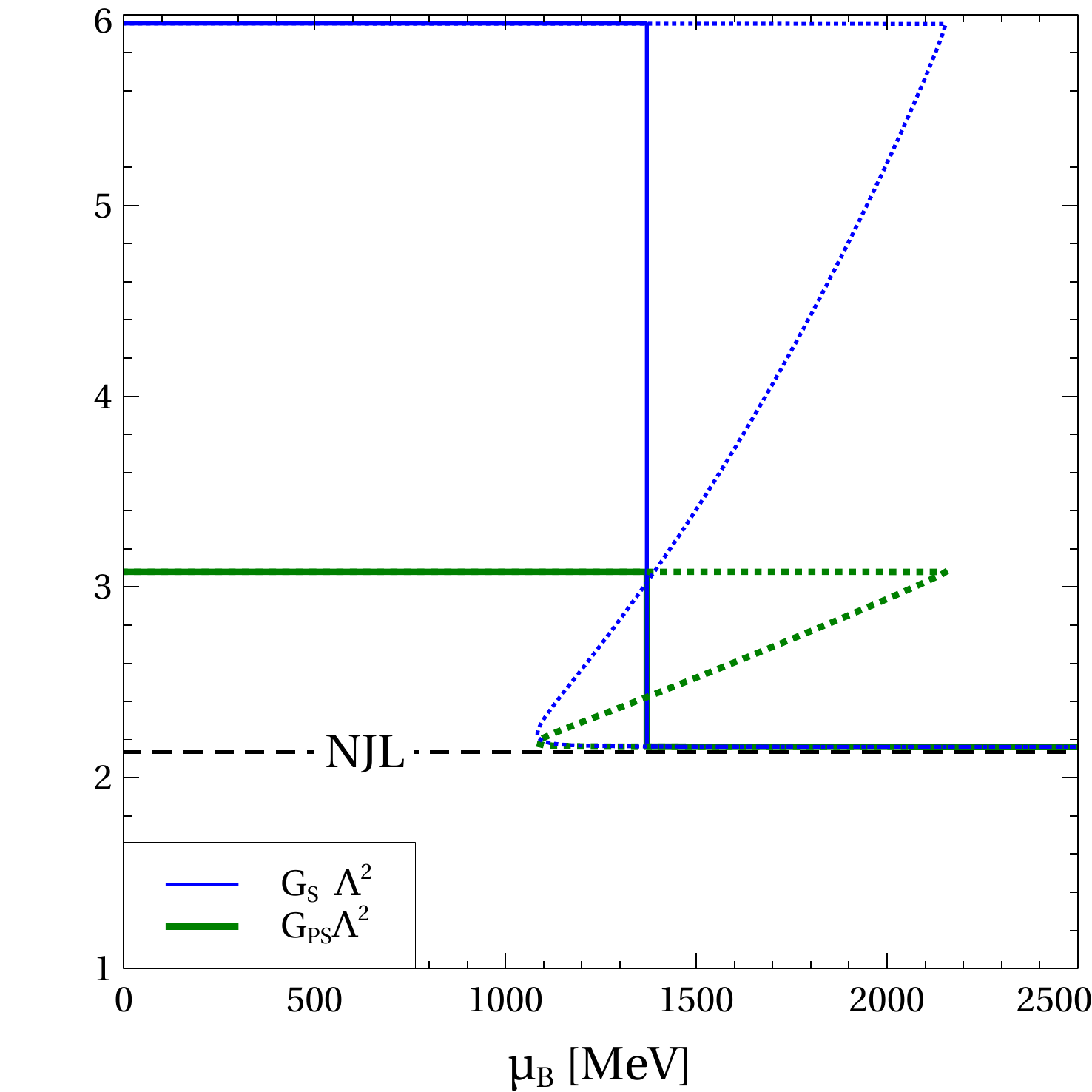}
\caption{Scaled effective scalar $G_S\Lambda^2$ (thin solid curves) and pseudoscalar $G_{PS}\Lambda^2$ (thick solid curve) couplings as functions of temperature $T$ at $\mu_B=0$ (upper panel) and baryonic chemical potential $\mu_B$ at $T=0$ (lower panel). Dotted curves on the lower panel demonstrate unstable parts removed by applying the Maxwell construction. Dashed line represents the NJL value $G_{NJL}\Lambda_{NJL}^2=2.14$ from Ref. \cite{Ratti:2005jh}. Calculations are performed for symmetric quark matter and $G_V=G_D=0$ while the rest of the model parameters have the values from Table \ref{table1}.}
\label{fig2}
\end{figure}

Eq. (\ref{XXXII}) provides $\hat m^*=\hat m-2G_{PS}\langle\overline{q}q\rangle$ with $G_{PS}$ defined by Eq. (\ref{V}). 
This relation is similar to the one of the NJL model \cite{Klevansky:1992qe}. 
For $\varkappa=1$ the two models coincide. 
The insignificant constant term $-D_0(1+\alpha)\langle\overline{q}q\rangle_0^2$ in the expression for the Lagrangian can be eliminated by $\alpha=-1$. 
In our approach, however, effective couplings are medium dependent. As is seen from Fig. \ref{fig2}, they decrease with temperature and chemical potential, which is in line with the weakening of the strong interaction due to the running of the QCD coupling \cite{Gross:1973id,Politzer:1973fx,Deur:2016tte}. 
It is worth mentioning, that a large value of $G_{PS}$ is another reason for high quark masses in the confining region.

Furthermore, contrary to the NJL model, the couplings in the scalar and pseudoscalar channels of the present model do not coincide in the general case, i.e. $G_S\neq G_{PS}$.
This signals the violation of the chiral symmetry on the level of the Lagrangian $\mathcal{L}^{(2)}$ and is a direct consequence of expanding $\mathcal{U}$ around the mean-field solution, which is known to break the chiral symmetry. 
As can be seen from Fig. \ref{fig2}, 
at certain values of $T$ and/or $\mu_B$, however, these couplings converge to the same constant value $G_\infty$ 
as a consequence of the dynamical restoration of the chiral symmetry. 
The value for $G_\infty$ is independent of the coupling constants in the vector and diquark channels. 
For the model parameters from Table \ref{table1} the model yields $G_\infty=6.6~{\rm GeV}^{-2}$.
It is remarkable that this asymptotic value of scalar and pseudoscalar couplings corresponds to
$G_\infty \Lambda^2 =2.16$, which is very close to the value $ G_{NJL} \Lambda^2_{NJL}=2.14$ obtained in the NJL model  \cite{Ratti:2005jh}. 

%

\section{Parametrization of the model}
\label{sec5}

\begin{table}[b]
\begin{tabular}{|c|c|c|c|c|c|c|c|c|c|c|c|}
\hline
  $m$ [MeV] & $\Lambda$ [MeV] & $\alpha$ & $D_0\Lambda^{-2}$  \\ \hline 
      4.2   &    573    &    1.43  &  1.39    \\ \hline \hline
 $M_\pi$ [MeV]& $F_\pi$ [MeV] & $M_\sigma$ [MeV] & $\langle\overline{l}l\rangle_0^{1/3}$ [MeV]\\ \hline
 140 & 92  &  980  & -267     \\ \hline
\end{tabular}
\caption{Parameters of the model used and resulting physical quantities.}
\label{table1}
\end{table}

\begin{figure}[t]
\includegraphics[width=0.9\columnwidth]{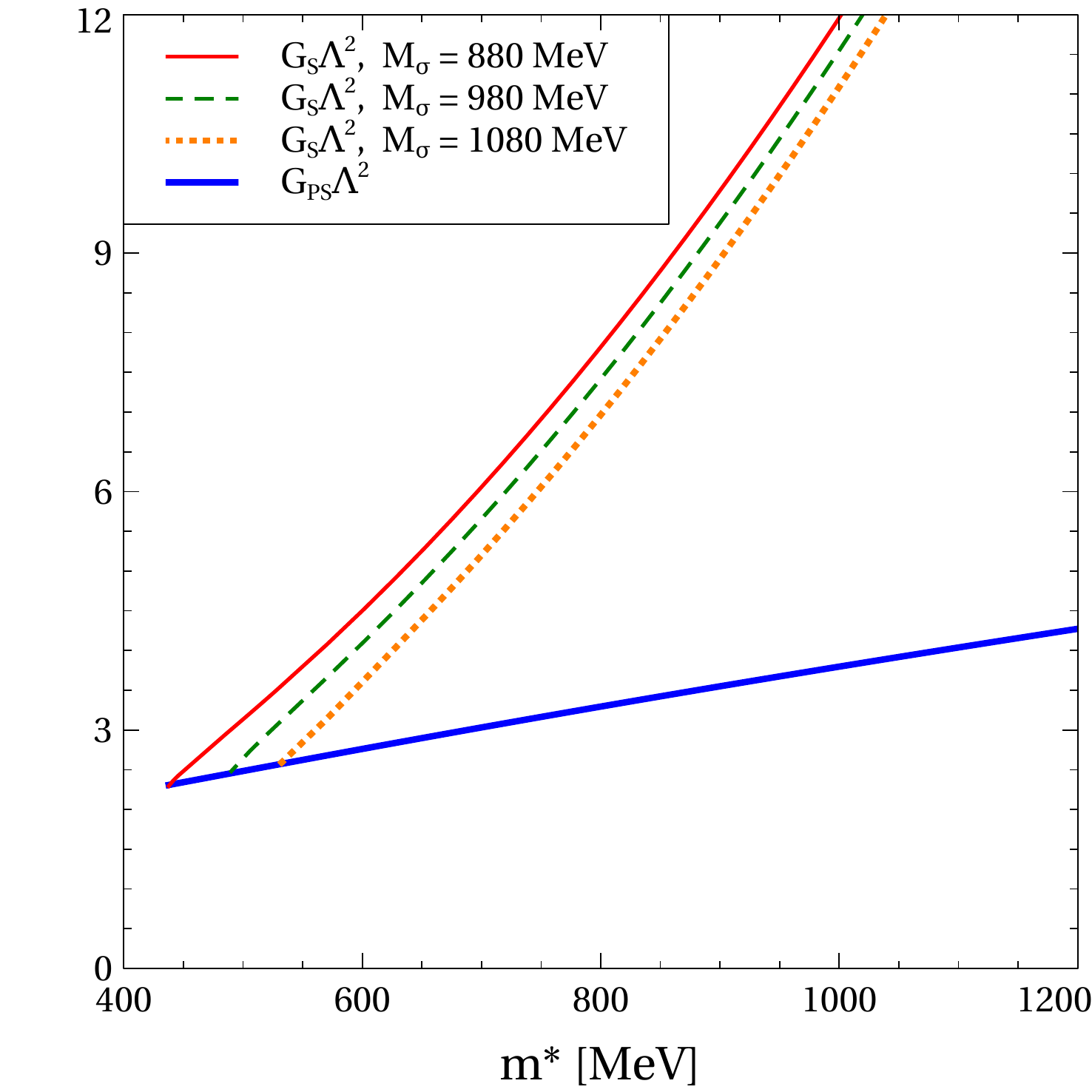}
\caption{Dimensionless scalar ($G_S\Lambda^2$, thin curves) and pseudoscalar ($G_{PS}\Lambda^2$, thick curve) couplings as functions of the effective quark mass $m^*$ calculated for the vacuums state.}
\label{fig3}
\end{figure}

The model has six independent parameters: the current quark mass $m$ which is degenerate for the two flavors, the parameters of the density functional, $\alpha$ and $D_0$, the momentum scale parameter $\Lambda$, the vector coupling $G_V$ and the diquark coupling $G_D$. 
The first four of them define the vacuum state of the model while the coupling constants $G_V$ and $G_D$ are free parameters determining its high-density behavior. 
The consideration of the present section is limited to the vacuum state. 
Most important for the parametrization of the model are the properties of the pseudoscalar meson, the pion. 
Namely its mass $M_\pi$ and decay constant $F_\pi$, with their well-known values given in Table \ref{table1}. 
The
vacuum value of the chiral condensate per flavor $\langle\overline{l}l\rangle_0={\langle\overline{q}q\rangle_0}/2$ is another important quantity. 
Predictions for $\langle\overline{q}q\rangle_0$ from chiral quark models of NJL type with a soft cutoff \cite{Grigorian:2006qe} 
like in Eq. (\ref{XXII})
are in tension with the result obtained from QCD sum rules at the renormalization scale of 1 GeV, which is  \cite{Jamin:2002ev} $|\langle\bar{l}l\rangle_0^{1/3}|\big|_{\rm 1~GeV}=242\pm 15$ MeV.
Therefore, we allow a deviation of this quantity from $\langle\overline{l}l\rangle_0\big|_{\rm 1~GeV}$. 
The scalar meson mass $M_\sigma$ is also considered. 
Despite possibly being the lightest scalar meson $f_0(500)$ has rather large width 500 - 1000 MeV \cite{PhysRevD.98.030001}, which makes its relevance to the vacuum phenomenology quite doubtful. 
Therefore, for our analysis we use the $f_0(980)$ meson with a width $\lesssim$ 100 MeV \cite{PhysRevD.98.030001}. 
Its mass is varied within the range from 880 MeV to 1080 MeV.
We use the following strategy. 
For a given value of the effective quark mass $m^*$ in the vacuum we fit $m$ and $\Lambda$ to $M_\pi$ and $F_\pi$. 
This allows us to find $\langle\overline{l}l\rangle_0$ as well as the pseudoscalar coupling $G_{PS}=(m-m^*)/(2\langle\overline{q}q\rangle_0)$.  
Having the above parameters fitted and varying $M_\sigma$ within the mentioned interval we can calculate the scalar coupling $G_S$ from Eq. (\ref{XXIX}). 
Fig. \ref{fig3} shows $G_{PS}$ and $G_{S}$ as functions of $m^*$. 
The calculations are performed for different $m^*$ which provide $G_S>G_{PS}$ and thus represent a range of physical values of the model parameters. 
Indeed, we can show by direct calculations that in the vacuum
\begin{eqnarray}
\label{XXXIV}
G_{S}-G_{PS}=\frac{2D_0}{9(1+\alpha)^{5/3}\langle\overline{q}q\rangle_0^{4/3}}>0.
\end{eqnarray}
This quantity vanishes only for $\alpha\rightarrow \infty$ since finite scalar and pseudoscalar couplings require $D_0\neq0$. 
Thus, $\alpha$ remains finite only 
when the effective quark mass $m^*$ exceeds some value, which mildly depends on the mass of the scalar mode. 
It is seen from Fig. \ref{fig3} that this value grows with $M_\sigma$.

\begin{figure}[t]
\includegraphics[width=0.9\columnwidth]{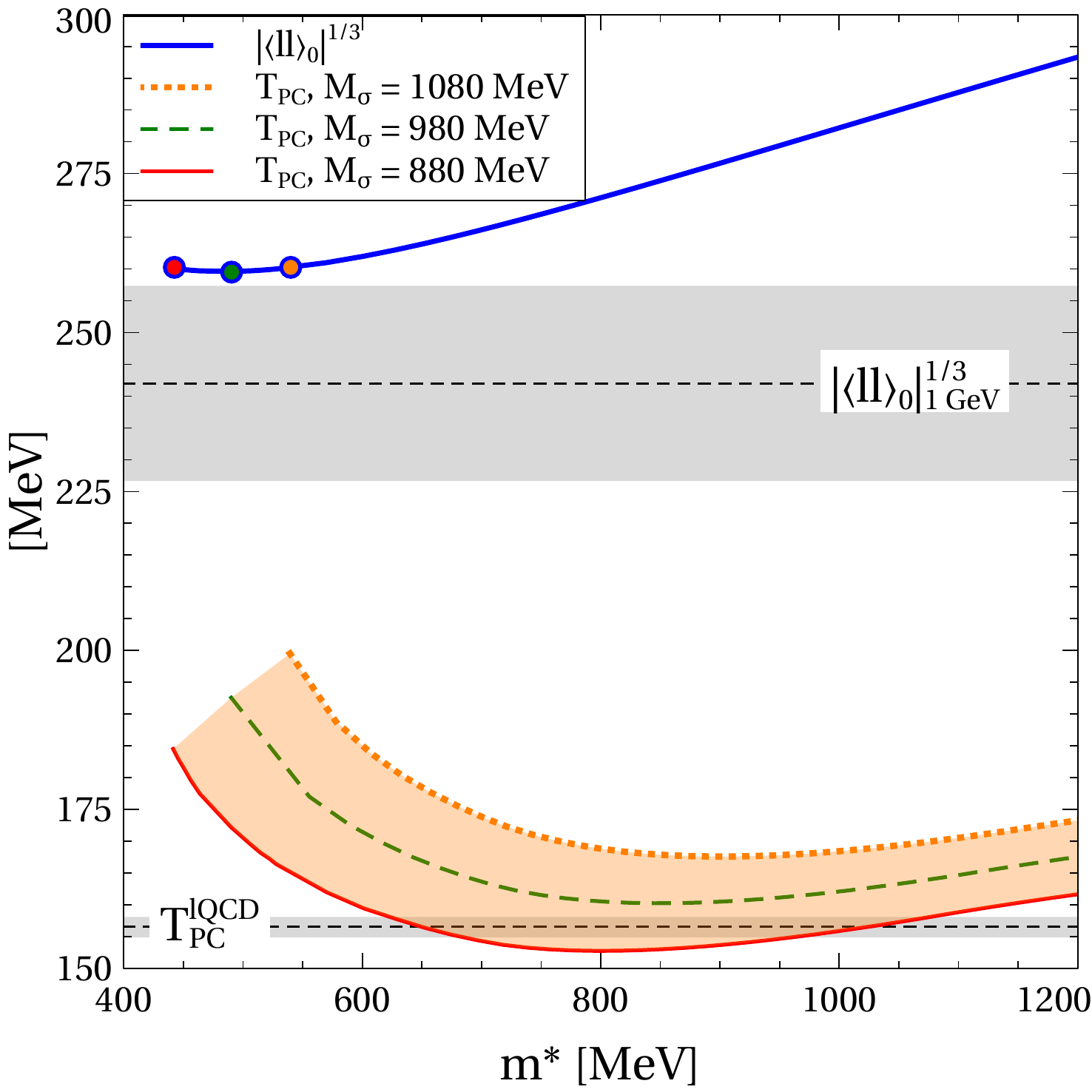}
\caption{Chiral condensate per flavor $|\langle\overline{l}l\rangle_0|^{1/3}$ (thick curve) and pseudocritical temperature $T_{PC}$ (thin curves) as functions of the the effective quark mass $m^*$ in the vacuum. Filled circles represent the low mass edges of the intervals where parameters of the model attain physical values (see description in the text). 
The orange shaded area covers the range of $T_{PC}$ generated by the considered vacuum values of $m^*$ and $M_{\sigma}$. Dashed horizontal lines and grey shaded areas represent $|\langle\overline{l}l\rangle_0^{\rm 1~GeV}|^{1/3}$ and $T_{PC}^{\rm lQCD}$ along with the corresponding error bars obtained with QCD sum rules and lattice QCD, respectively. }
\label{fig4}
\end{figure}

For a given set of the model parameters the pseudocritical temperature $T_{PC}$ can be found according to its definition as the peak position of the chiral susceptibility, which is defined as the temperature derivative of the chiral condensate at vanishing chemical potential. 
This quantity depends on both effective quark mass $m^*$ and mass of scalar meson $M_\sigma$ in the vacuum. 
Note  that $T_{PC}$ calculated in such a way is just a preliminary result since it is limited to the approximation that is employed when calculating the chiral condensate. When the mean-field approximation is used then it does not account for the hadronic correlations of quarks. 
Such correlations are known to contribute to the melting of the chiral condensate in the pseudocritical region \cite{Hufner:1994ma,Zhuang:1994dw} and lead to a lowering of the pseudocritical temperature. 
Therefore, we require our $T_{PC}$ to exceed the lattice QCD result $T_{PC}^{\rm lQCD}=156.5\pm  1.5$ MeV \cite{HotQCD:2018pds}. 
Fig. \ref{fig4} shows $|\langle\overline{l}l\rangle_0|^{1/3}$ and $T_{PC}$ as functions of $m^*$. 
At large effective quark mass the chiral condensate is too high at any value of the scalar meson mass. 
Therefore, we require $m^* \lesssim 800$ MeV. 
At the same time we prefer to have $m^*$ high enough to provide an efficient suppression of quarks in the confinement region. 
This excludes low $M_\sigma$, since in this case either $T_{PC}<T_{PC}^{\rm lQCD}$ or $m^*$ is not high enough. 
On the other hand, for $M_\sigma=1080$ MeV the pseudocritical temperature $T_{PC}$ of the model exceeds the one from lattice QCD $T_{PC}^{\rm lQCD}$ by more than 10\%. 
Due to this we use the central value of $M_\sigma=980$ MeV. At $m^*=718$ MeV this value of the scalar meson mass leads to a reasonable value of $T_{PC}=163$ MeV and $|\langle\overline{l}l\rangle_0|^{1/3}=267$ MeV compatible with results of other chiral models \cite{Grigorian:2006qe}. The corresponding values of the model parameters are summarized in Table \ref{table1}.
It is worth mentioning that a relatively large value of the vacuum effective quark mass in our approach is quite in line with the results of an analysis of the data on the lattice QCD thermodynamics performed with the quark quasiparticle model yielding $m^*_0= 610 - 950$ MeV \cite{Plumari:2011mk} or even $m_0^*>1$ GeV \cite{Bozek:1998dj}.


First principle information about vector and diquark couplings is absent at the moment. 
The argumentation based on a Fierz transformation of the (massive) vector boson exchange interaction model which yields $G_V=0.5~G_S$ and $G_D=0.75~G_S$ \cite{Buballa:2003qv} is not directly applicable to the present model since its (pseudo)scalar couplings are not constant. 
Therefore, $G_V$ and $G_D$ are treated as free parameters. 
We use the vacuum value of the scalar coupling $G_{S0}=18.1~{\rm GeV}^{-2}$ in order to parameterize them by $\eta_V\equiv\frac{G_V}{G_{S0}}$ and $\eta_D\equiv\frac{G_D}{G_{S0}}$. 
The model parameterizations considered below are labeled with a pair of numbers corresponding to scaled couplings $\eta_V$ and $\eta_D$. For example, $(0.2,1.5)$ stands for a quark matter EoS obtained for $\eta_V=0.2$, $\eta_D=1.5$ and other parameters from Table \ref{table1}.  

\section{Cold quark matter}
\label{sec6}

\begin{figure}[t]
\includegraphics[width=0.9\columnwidth]{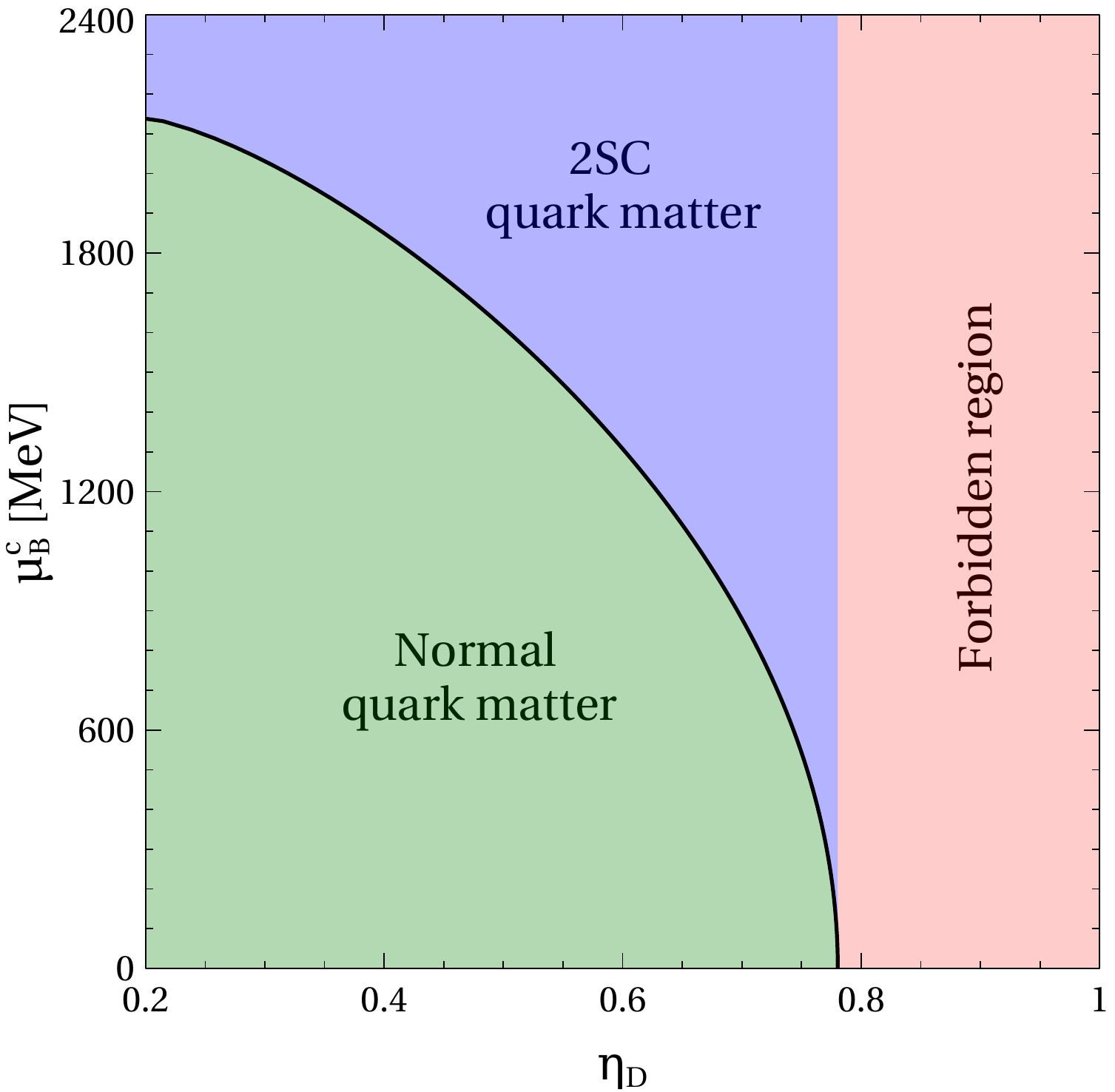}
\caption{Baryonic chemical potential of the 2SC phase onset $\mu_B^c$ as a function of the scaled diquark coupling $\eta_D$. The calculations are performed for symmetric quark matter and $\eta_V=0$.}
\label{fig5}
\end{figure}

From now on we consider the case of zero temperature, which is of interest for the astrophysical applications. The first question we address is related to onset of the 2SC phase. It occurs when two solutions of Eq. (\ref{XXI}) coincide provided by 
\begin{eqnarray}
\label{XXXV}
\frac{\partial^2\Omega}{\partial\Delta^2}\biggl|_{\Delta=0}=0.
\end{eqnarray}
Solving this equation with respect to baryonic chemical potential yields its critical value $\mu_B^c$ corresponding to the onset of color superconductivity. Fig. \ref{fig5} shows it as a function of the scaled diquark coupling for the case of symmetric quark matter with $\mu_Q=0$. At chemical potentials below this critical value quark matter exists in a normal state, while $\mu_B>\mu_B^c$ corresponds to 2SC phase. Larger diquark couplings lead to stronger pairing among quarks and, consequently, yield smaller $\mu_B^c$. 
At a certain value of the diquark coupling $\eta_D^*$ the critical chemical potential vanishes, meaning that color superconductivity is supported by the vacuum. 
Numerical analysis yields $\eta_D^*=0.78$.  
We note that this result is not affected by the value of $\eta_V$ since vector repulsion is irrelevant in the vacuum.  In order to prevent formation of color superconductivity in the vacuum we set the constraint $\eta_D<\eta_D^*$. 
It is interesting to compare $\eta_D^*$ to the maximal value of the diquark coupling for which color superconductivity in the vacuum is still excluded. 
In the NJL model this is $({3}{m^*}/[2(m^*-m)]\simeq{3}/{2}$ \cite{Zablocki:2009ds}.
Our value of $\eta_D^*$ is about twice smaller than the NJL one. This difference is due to the adopted definition of $\eta_D$. 
It can be exactly compensated by rescaling $\eta_D$ by the factor ${G_{S0}}/{G_{PS0}}$.
Indeed, in the vacuum $f_{{\bf k}fc}^\pm=0$ and Eq. (\ref{XXII}) at vanishing temperature, chemical potential and diquark pairing gap yields
\begin{eqnarray}
\label{XXXVI}
\langle\overline{q}q\rangle=-6\sum_f\int\frac{d{\bf k}}{(2\pi)^3}g_{\bf k}\frac{m^*}{\epsilon_{{\bf k}f}}.
\end{eqnarray}
Here and all along the derivation of $\eta_D^*$ we omit the subscript $``0"$ where it is applicable for the sake of shortening notations. Using this relation we obtain by a straightforward calculation
\begin{eqnarray}
\frac{\partial^2\Omega}{\partial\Delta^2}\biggl|_{\Delta=0}&=&\frac{1}{2G_D}-4\sum_f\int\frac{d{\bf k}}{(2\pi)^3}\frac{g_{\bf k}}{\epsilon_{{\bf k}f}}\nonumber\\
\label{XXXVII}
&=&\frac{1}{2G_D}+\frac{2\langle{q}q\rangle}{3m^*}=0,
\end{eqnarray}
This equation can be solved with respect to $G_D$. The corresponding solution combined with $m^*=m-2G_{PS}\langle\overline{q}q\rangle$ and the definition of $\eta_D$ yields
\begin{eqnarray}
\label{XXXVIII}
\eta_D^*=\frac{3}{2}\frac{m^*}{m^*-m}\frac{G_{PS}}{G_{S}}.
\end{eqnarray}
This expression coincides with the one of the NJL model only if $G_{S}=G_{PS}$, which is violated by the vacuum solution of the present model. 
At the same time, we would like to stress that the absolute value of $G_D^*=\eta_D^* G_{S0}$ is about twice larger compared to the maximal value of the diquark coupling in the NJL model. 
This provides the present approach with a unique possibility to access the region of strong diquark pairing that is interesting from the phenomenological point of view.

\begin{figure}[t]
\includegraphics[width=0.9\columnwidth]{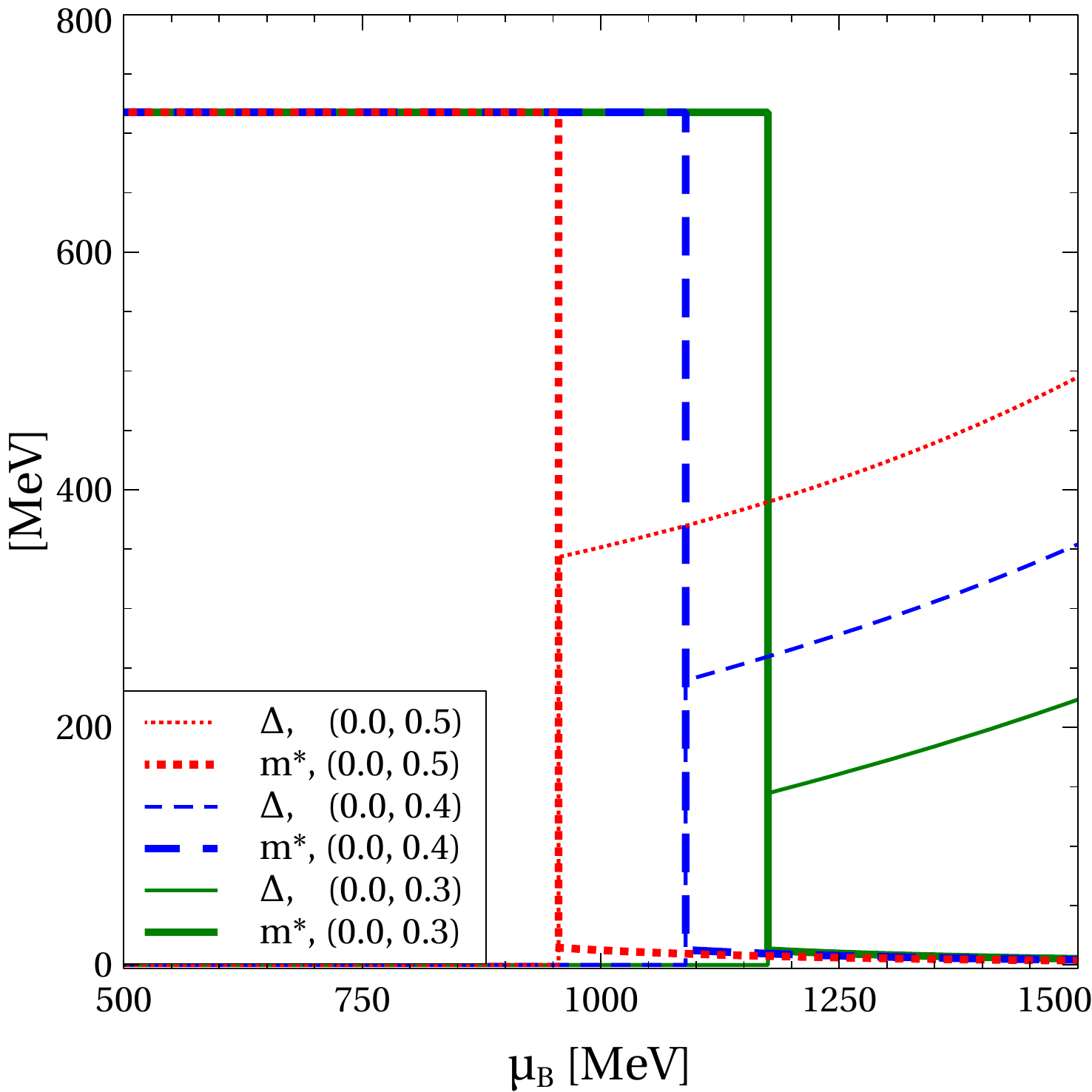}
\caption{Effective quark mass $m^*$ (thick curves) and diquark pairing gap $\Delta$ (thin curves) as functions of baryonic chemical potential $\mu_B$ calculated for electrically neutral $\beta$-equilibrated matter and $\eta_V=0$.}
\label{fig6}
\end{figure}
%

\section{Compact stars with quark cores}
\label{sec7}

In order to model compact stars as astrophysical objects 
observable as pulsars, we need the $T=0$ EoS under the conditions of electric neutrality and $\beta$-equilibrium. 
The former is provided by a proper amount of non-interacting electrons with the chemical potential $\mu_e$ and the negligible mass $m_e=0.511$ MeV, while the latter is ensured by
$\mu_e=\mu_d-\mu_u=-\mu_Q$. 
The effective quark mass and the diquark pairing gap calculated under these conditions are shown as functions of the baryonic chemical potential in Fig. \ref{fig6}. 
At small $\mu_B$ chiral symmetry is broken, which leads to high quark masses and the absence of color superconductivity, as is indicated by a vanishing pairing gap. 
At a certain value of the chemical potential $\mu_B$ the effective mass $m^*$ drops and the diquark pairing gap $\Delta$ attains a finite value. 
The discontinuous character of this change signals a first order phase transition. 
Further increase of $\mu_B$ leads to a relatively slow decrease of effective mass and a further growth of the pairing gap. 
At asymptotically high densities holds $m^*\propto(\Delta\sum_f\mu_f^*)^{-1}$ and $\Delta\propto\sum_f\mu_f^{*3}$, see the Appendix B for the details. 
In full agreement with the discussion of $\mu_B^c$, an increase of the diquark coupling lowers the chemical potential of the chiral symmetry restoration and color superconductivity onset. This picture holds for any value of the vector coupling, which simply renormalizes $\mu_B$.

\begin{figure}[t]
\includegraphics[width=0.9\columnwidth]{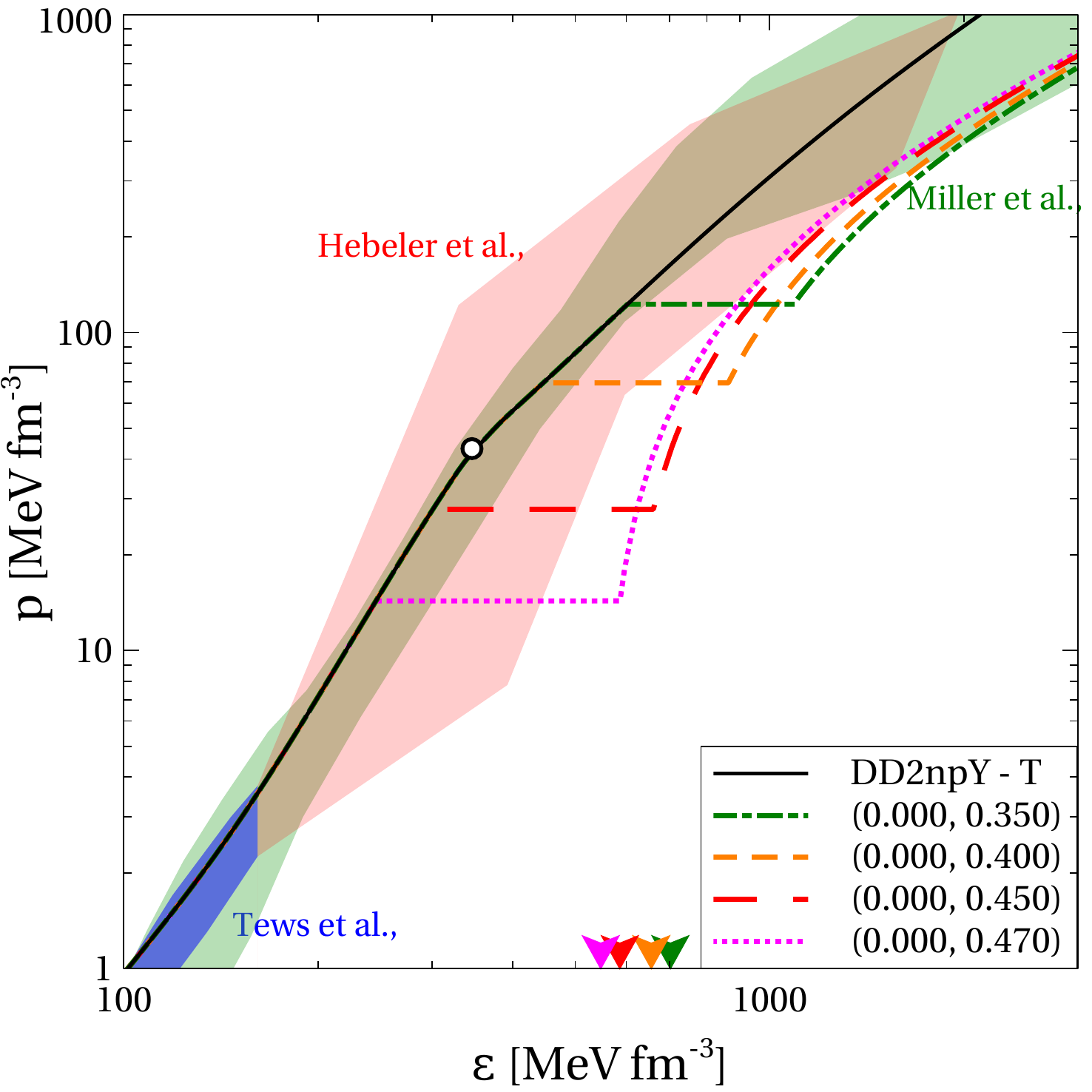}\\
\includegraphics[width=0.9\columnwidth]{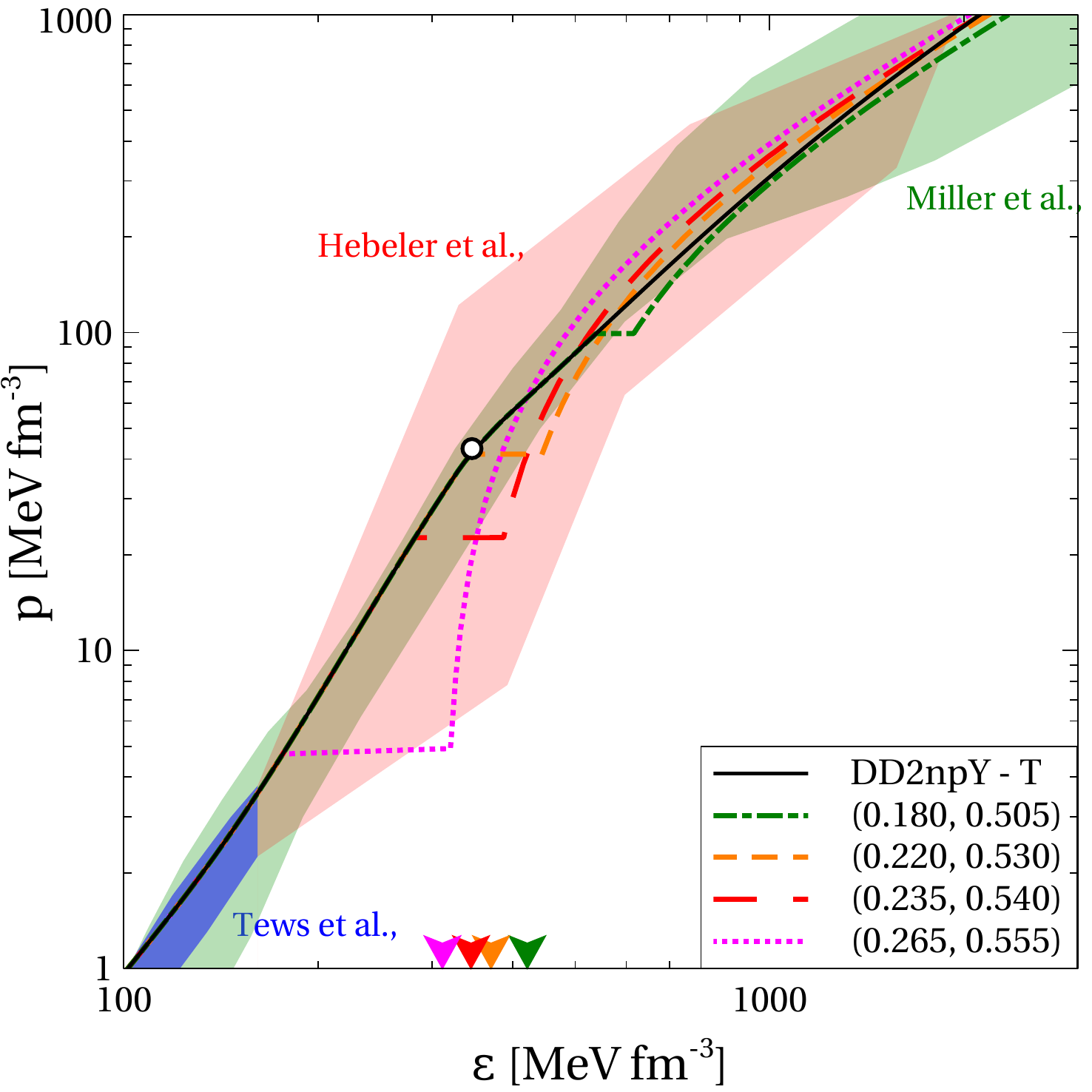}
\caption{Hybrid EoS of cold electrically neutral $\beta$-equilibrated quark-hadron matter in the plane of energy density $\varepsilon$ and pressure $p$. Empty circles on the hadronic curves indicate the hyperon onset. The calculations are performed for $\eta_V=0$ (upper panel) and $\eta_V\neq 0$ (lower panel). The nuclear matter constraints represented by the shaded areas are discussed in the text.
Arrows indicate the energy densities of the 2SC phase onset in pure quark matter EoS, being a part of the hybrid EoS curve of the corresponding color. These energy densities are 706 $\rm MeV~fm^{-3}$, 657 $\rm MeV~fm^{-3}$, 582 $\rm MeV~fm^{-3}$ and 546 $\rm MeV~fm^{-3}$ for the upper panel and 422 $\rm MeV~fm^{-3}$, 370 $\rm MeV~fm^{-3}$, 347 $\rm MeV~fm^{-3}$ and 313 $\rm MeV~fm^{-3}$ for the lower one.}
\label{fig7}
\end{figure}

Having vector field, diquark pairing gap and chiral condensate self consistently found from Eqs. (\ref{XX}) - (\ref{XXII}), we can construct the EoS of electrically neutral $\beta$-equilibrated quark matter. 
The hybrid EoS of stellar matter is obtained by merging this quark matter EoS to the hadronic one via the Maxwell construction corresponding to the first order phase transition. 
The position of this transition is defined by requiring equal pressures of quark and hadron phases at the same value of baryonic chemical potential. 
In this case baryon number and energy densities experience a discontinuous change along the transition. Its details are sensitive to the values of vector and diquark couplings. 
In this work hadronic matter is described by the DD2 EoS in its formulation and parametrization from Ref. \cite{Typel:2009sy}. 
We use its version with hyperons DD2npY-T \cite{Shahrbaf:2022upc} that is supplemented with a crust EoS. 
This relativistic density functional model introduces interaction between nucleons via meson exchange. 
Besides reproducing the properties of normal nuclear matter, the DD2 EoS is in a good agreement with the low density calculations of the chiral EFT approach \cite{Kruger:2013kua}. 
The high density behavior of the DD2npY-T model can be improved with the excluded volume correction \cite{Typel:2016srf}. 
Within our approach, however, this correction is neglected since at high densities hadronic degrees of freedom are replaced by the quark ones.  Fig. \ref{fig7} shows the hybrid quark-hadron EoS in the plane of energy density and pressure.
Notably, the onset of the 2SC phase in quark matter occurs at densities typical for the hadron branch of the hybrid EoS. Therefore, the quark matter branch is already color superconducting. Values of densities of the 2SC phase onset in quark matter depend on the vector and diquark couplings and are given in the caption of Fig. \ref{fig7}.
As is seen from its upper panel, larger diquark couplings lead to an earlier onset of quark matter. 
The width of the mixed phase region, however, depends on $\eta_D$
only weakly. 
We would like to stress, that the later the quark matter onset, the softer is its EoS. 
At vanishing $\eta_V$, the quark part of the EoS is too soft and inconsistent with constraints obtained within the multipolytrope analysis of the observational data of PSR J1614+2230 \cite{Hebeler:2013nza} and PSR J0740+6620 \cite{Miller:2021qha}.
This redundant softening can be compensated by the vector mean field repulsion in quark matter. 
The corresponding effect is seen on the lower panel of Fig. \ref{fig7}. 
It is remarkable, that the stiffening of the quark EoS can be so strong that the corresponding pressure can sizably exceed the hadronic one. 
This yields a positive feedback to the problem of reaching the two solar mass limit by compact stars. 
The corresponding sets of the vector and diquark couplings lead to an early onset of quark matter with the densities either lower or very close to the one at which hyperons appear in the purely hadronic matter (white circles on Fig. \ref{fig7}). 
As a result, within our approach the hyperon puzzle \cite{Chatterjee:2015pua} is resolved by early onset of quark matter, which already has been proposed as a possible solution, see \cite{Shahrbaf:2019vtf} for a recent work and more literature cited therein. 
As a generic feature of the present model we can conclude that the stiffness of the quark matter EoS is controlled by the vector coupling, while the diquark one regulates the quark matter onset density. 
Our hybrid EoS is in a good agreement with the constraints from Ref. \cite{Hebeler:2013nza}. 
A similar constraint from Ref. \cite{Miller:2021qha} is fulfilled by the present approach only at high densities. Some discrepancy at small densities is caused by the strong phase transition from hadrons to quarks. 
This allows us to conclude about the limited ability of multipolytropic EoS to catch a strong first order phase transition from hadronic to quark matter. 
It is worth mentioning that a nonvanishing $\eta_V$ narrows the mixed phase region when compared to the case of $\eta_V=0$. 

\begin{figure}[t]
\includegraphics[width=0.9\columnwidth]{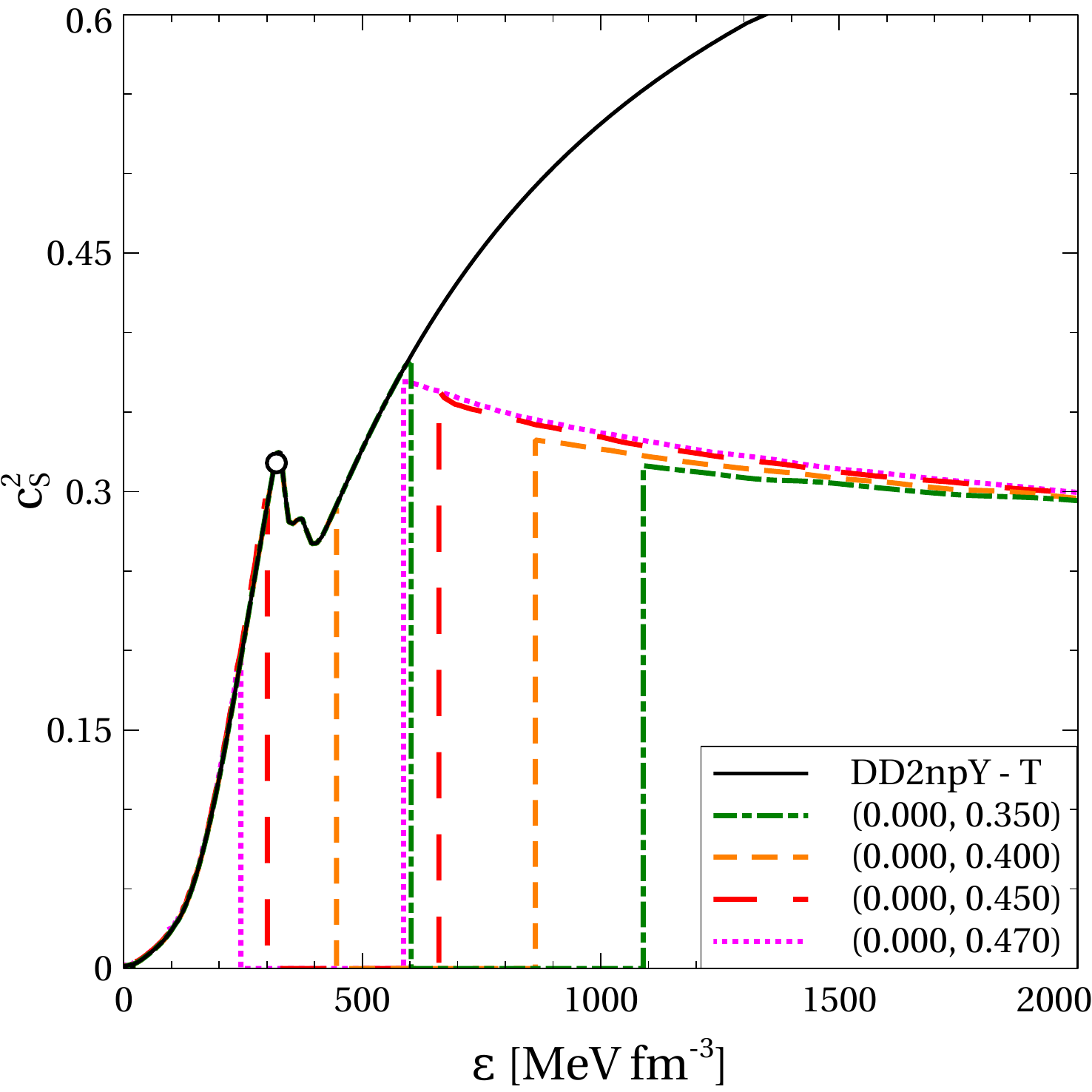}\\
\includegraphics[width=0.9\columnwidth]{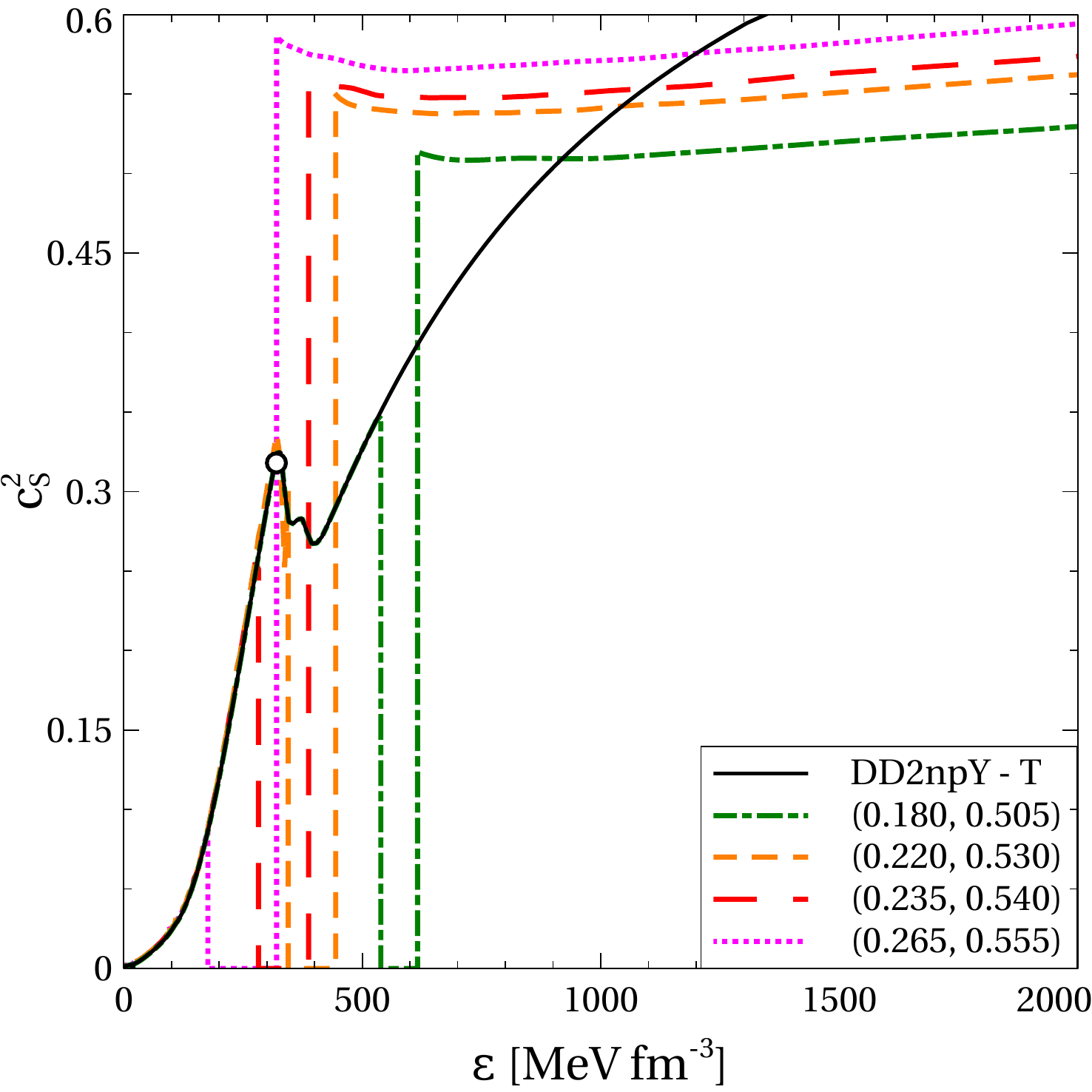}
\caption{Squared speed of sound $c_S^2$ of stellar matter as a function of energy density $\varepsilon$ calculated with the quark-hadron EoS presented in Fig. \ref{fig7}. Empty circles on the hadronic curves indicate the hyperon onset.}
\label{fig8}
\end{figure}

The stiffness of an EoS can be quantified by the (squared) speed of sound $c_S^2={d p}/{d\varepsilon}$. 
As is shown in the Appendix B, at high densities and $\eta_V=0$, color superconducting quark matter is characterized by  $c_S^2\rightarrow {1}/{5}$, while a finite $\eta_V$ yields $c_S^2\rightarrow 1$. 
Fig. \ref{fig8} shows this quantity as a function of energy density. In the absence of vector repulsion and at densities typical for the compact star interiors, the squared speed of sound of quark matter varies within the range 0.3-0.35. 

At vanishing vector coupling only the second of these scenarios can be realized if quark matter onsets before two saturation densities provided by $\eta_D>0.45$. However, this range of values of the diquark coupling is inconsistent with the constraints on the stellar matter EoS from Refs. \cite{Hebeler:2013nza,Miller:2021qha}. 
This discrepancy can be resolved by applying finite vector couplings. 
The lower panel of Fig. \ref{fig8} indicates, that the speed of sound increases when the vector coupling is increased, reflecting a stiffening of the quark matter EoS. 
It is remarkable that for a given value of $\eta_V$ the speed of sound of quark matter remains almost constant in the density range typical for compact stars. 
For $\epsilon\le 2000~{\rm MeV~fm}^{-3}$ the corresponding variation does not exceed few percent. 
In other words, our approach supports the constant speed of sound (CSS) parameterization \cite{Alford:2013aca,Zdunik:2012dj}, which is widely applied to model quark matter, to classify compact stars with quark cores \cite{Blaschke:2020vuy,Miao:2020yjk} and to study the third and fourth stellar families \cite{Paschalidis:2017qmb,Alford:2017qgh,Li:2019fqe}.
It has been found recently that the EoS of color superconducting nonlocal chiral quark models of the NJL type results in an almost constant sound speed with $c_S^2$ in the narrow range between 
0.45 and 0.54 \cite{Antic:2021zbn,Contrera:2022tqh}.

\begin{figure}[t]
\includegraphics[width=0.9\columnwidth]{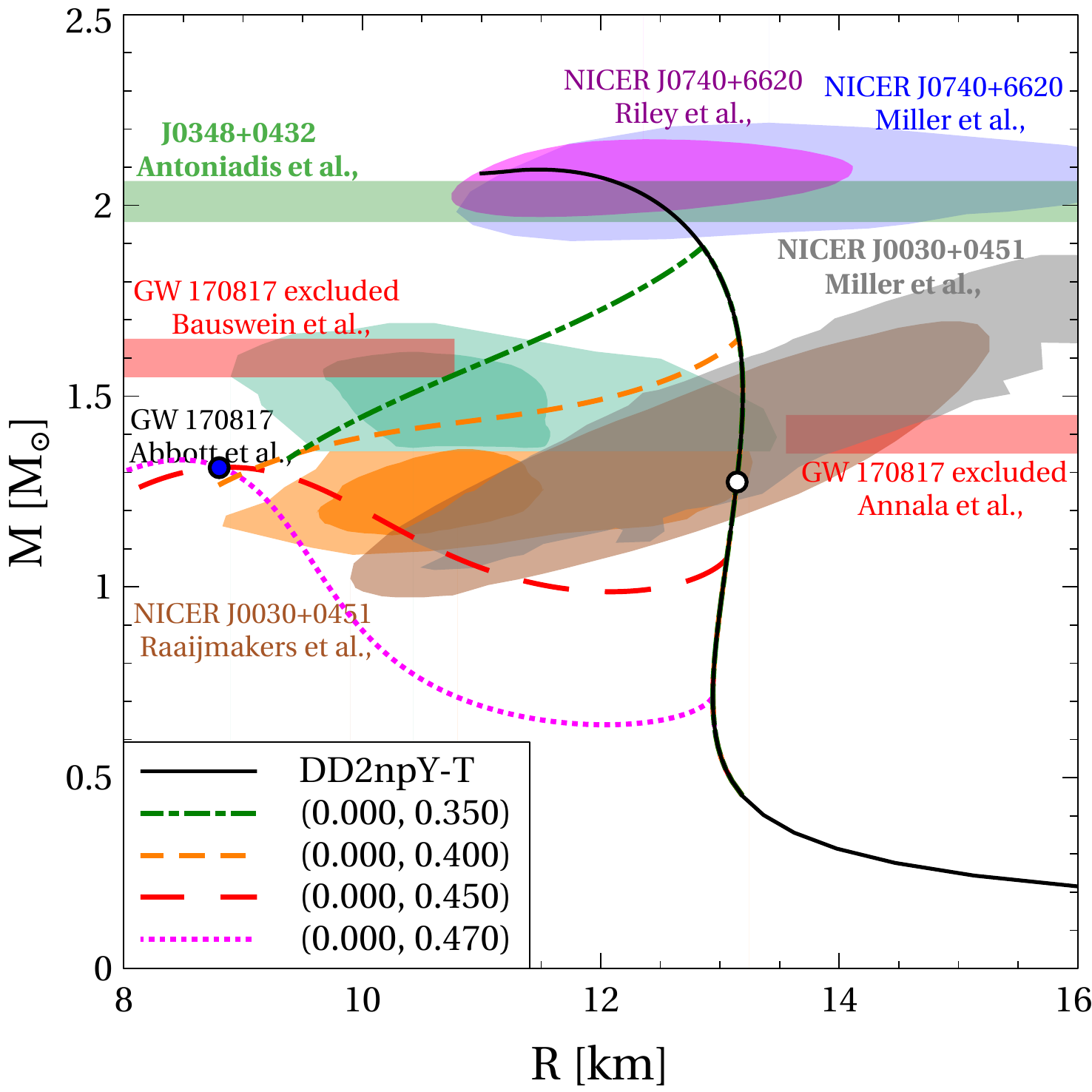}\\
\includegraphics[width=0.9\columnwidth]{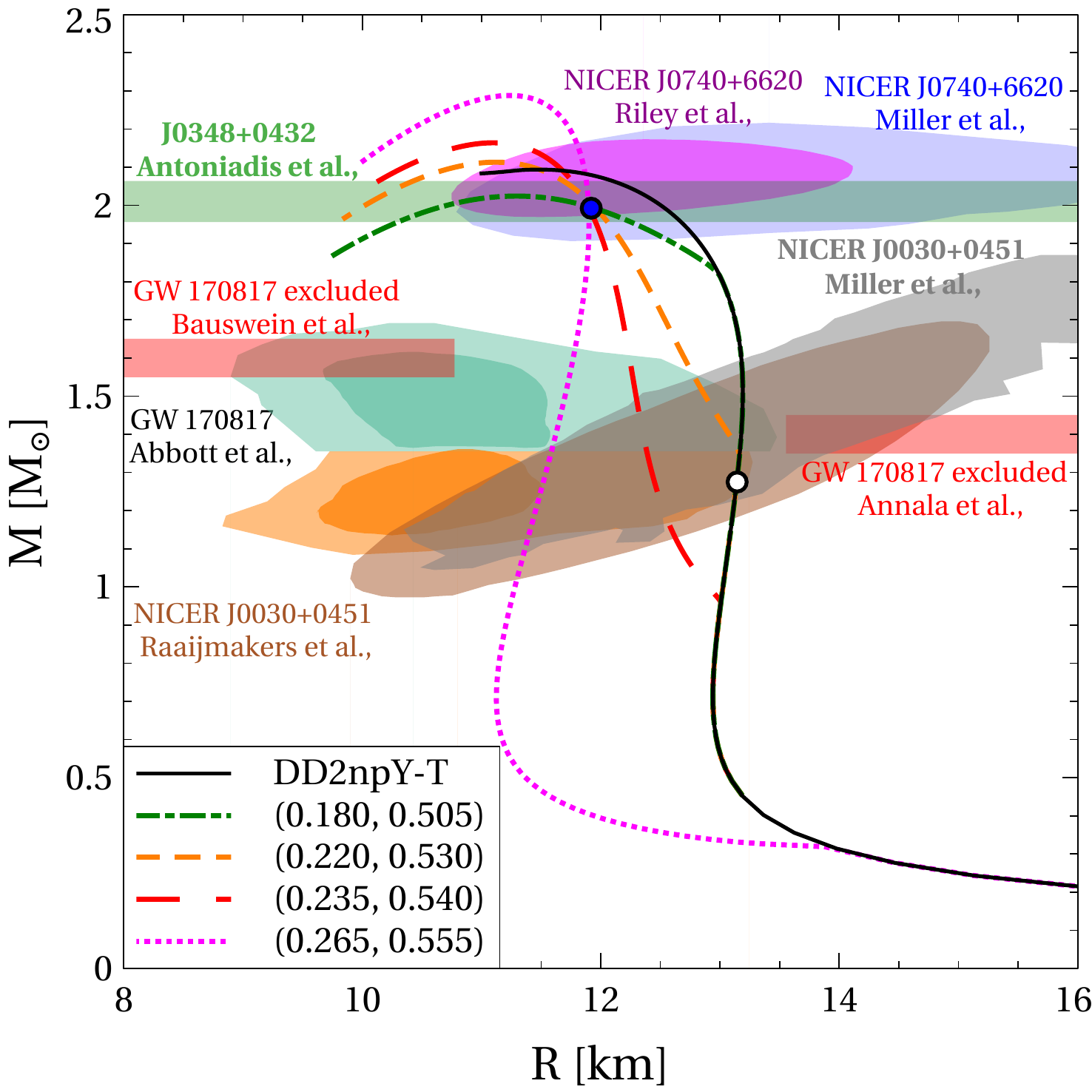}
\caption{Mass-radius relation of hybrid neutron stars with the quark-hadron EoS presented on Fig. \ref{fig7}. 
An empty circle on the hadronic curves indicates the hyperon onset. The blue filled circles represent the special points with the mass $\rm M_{SP}$ found according to the fitting procedure described in the text. The astrophysical constraints depicted by the colored bends and shaded areas are discussed in the text.}
\label{fig9}
\end{figure}

We apply the family of developed EoS to model compact stars with quark cores by solving the problem of relativistic hydrostatic equilibrium represented by the TOV equations with the corresponding boundary conditions. 
Fig. \ref{fig9} shows the resulting relation between the stellar radius $\rm R$ and mass $\rm M$. 
Our calculations are confronted with the constraint on the lower limit of the TOV maximum mass given by the mass $2.01^{+0.04}_{+0.04}~\rm M_\odot$ measured in a binary system of the pulsar PSR J0348+0432 and its white dwarf companion \cite{Antoniadis:2013pzd}. 
In the high and intermediate mass regions we also utilize constraints from the Bayesian analysis of the observational data from PSR J0740+6620 \cite{Riley:2021pdl,Miller:2021qha} and PSR J0030+0451 \cite{Riley:2019yda,Raaijmakers:2019qny}, respectively. A very important information about the intermediate mass region of the $\rm M$ - $\rm R$ diagram comes from the analysis of the gravitational wave signal produced by the merger GW170817 \cite{LIGOScientific:2018cki}. 
It constrains masses and radii of the low and high mass objects in the binary as represented in Fig. \ref{fig9} by the orange and cyan shaded areas. 
The darker and lighter of them correspond to $1\sigma$ and $2\sigma$ confidence intervals. 
Finally, based on the analysis of the same gravitational wave signal we limit the star radius at $1.6~\rm M_\odot$  from below by $R_{1.6}\ge 10.68$ km \cite{Bauswein:2017vtn} and at $1.4~\rm M_\odot$ from above by $R_{1.4}\le 13.6$ km \cite{Annala:2017llu}.

Regardless the values of $\eta_D$ and $\eta_V$, hybrid neutron stars with quark cores 
that are obtained by a Maxwell construction
are more compact compared to their purely hadronic counterparts of the same mass. 
This effect is especially attractive from the phenomenological point of view 
in the typical stellar mass range of about $1.4~{\rm M}_\odot$, where the recent observational data (in particular from GW170817  \cite{LIGOScientific:2018cki}) constrain the stellar radii to quite small values \cite{Capano:2019eae,Dietrich:2020efo}. 
Since our baseline hadronic EoS model DD2npY-T alone would not fulfill these new radius constraints, this explains why the phenomenologically interesting scenarios correspond to 
$\rm M_{onset}\lesssim 1.0~M_\odot$. 
The realization of such scenarios requires a relatively large $\eta_D$ 
because that would lead to an early onset of deconfinement. 
For the case of the three-flavor color superconducting NJL model
see \cite{Klahn:2006iw,Klahn:2013kga}. 
For the present model, this fact is illustrated in 
the upper panel of Fig. \ref{fig9}, where an increase of the diquark coupling dramatically lowers the onset mass of quark matter. 

However, at vanishing vector coupling, 
the model does not support the existence of stable hybrid stellar configurations with quark matter cores with masses above $2\rm M_\odot$. 
Moreover, in this case stable hybrid star branches the mass-radius diagram occur only if 
$\eta_D>0.43$.

We note that in this case, the remarkable feature of a third family of hybrid stars is realised 
\cite{Gerlach:1968zz,Alvarez-Castillo:2018pve,Blaschke:2020qqj}
because the stable second family branch of pure neutron stars is separated from the branch of stable hybrid stars by a sequence of unstable stars that is characterised by 
$\partial \rm M/\partial \rm \varepsilon_c < 0$ with $\varepsilon_c$ being energy density in the center of the star.
The existence of a third family is necessarily connected to the phenomenon of mass twin stars \cite{Glendenning:1998ag,Benic:2013eqa,Alvarez-Castillo:2014dva,Zacchi:2016tjw,Alvarez-Castillo:2017qki,Alford:2017qgh,Montana:2018bkb,Blaschke:2019tbh} the detection of which would be an observational proof of the existence of a strong phase transition in dense neutron star matter.

The problem with a too low maximum mass 
can be overcome with adopting finite values of the vector coupling. 
Increasing $\eta_V$ stabilizes quark branch of the mass-radius diagram and causes a positive feedback to the problem of reaching the two solar mass limit. 
However, the increase of $\eta_V$ moves $\rm M_{onset}$ towards higher stellar masses, which requires simultaneous increase of $\eta_D$ in order to keep $\rm M_{onset}<1.4~M_\odot$. 
We note that this requirement is set in order to obtain the necessary softening of the hybrid EoS for fulfilling the tidal deformability constraint formulated for $1.4~M_\odot$ stars, which is not fulfilled by the DD2npY - T EoS itself.
For example, at $\eta_D=0.530$ the maximum mass  $\rm M_{max}$ of the stable hybrid star sequence reaches $2.01~{\rm M_\odot}$ for $\eta_V=0.172$, while marginal agreement with the PSR J0740+6620 observational data is reached at $\eta_V=0.177$. 
Further increase of the vector coupling allows the present model to fulfil the above-mentioned astrophysical constraints. 
It is also interesting to note that at a certain value of $\eta_V$ defined by the value of $\eta_D$, the hybrid star branch of the mass-radius curve intersects the stable part of the branch corresponding to purely hadronic EoS models of neutron star matter. 
For $\eta_D=0.530$ this happens at $\eta_V=0.216$. 
In other words, at sufficiently large vector and diquark couplings, the existence of quark matter cores in the compact star interiors allows them to break through to the region of the mass-radius diagram inaccessible for purely hadronic stars due to the softening of their EoS caused by the appearance of hyperons. 
This possibility is important for compact star phenomenology since it might provide a hybrid neutron star explanation of the nature of the $2.6~\rm M_\odot$ companion of the $23~\rm M_\odot$ black hole recently reported in the merger GW190814 \cite{LIGOScientific:2020zkf}. 
We discuss this issue below.

High values of the vector and diquark couplings also allow the present model to achieve simultaneous agreement with the high and intermediate mass constraints coming from the NICER data on PSR J0740+6620 and from the gravitational wave signal of GW170817, which is a challenging task for most of the other approaches.

Fig. \ref{fig9} demonstrates a special aspect of star sequences in the mass-radius diagram that occurs only for hybrid stars, the so-called "special point". 
Strictly speaking, it is not a point but a narrow region of intersection of hybrid star sequences, which was first reported and analysed in Ref. \cite{Yudin:2014mla}. 
A thorough of its invariance properties has been given recently in Refs. \cite{Cierniak:2020eyh,Blaschke:2020vuy,Cierniak:2021knt,Cierniak:2021vlf}. 
The remarkable feature of the insensitivity of the special point location to the details of hadronic EoS and the construction of the phase transition, makes it a universal tool 
to study the properties of quark cores in compact stars. Within the CSS parameterization this special point is insensitive to constant bag pressure controlling the quark matter onset density. In our approach the same phenomenological role is played by the diquark coupling. This explains 
the occurrence of the special point when varying just $\eta_D$
in the upper panel of Fig. \ref{fig9}.
Its mass $M_{\rm SP}$ was found to obey empirical relation 
\begin{eqnarray}
\label{XXXIX}
\rm M_{max}=M_{SP}+\delta(M_{SP}-M_{onset})
\end{eqnarray}
with $\delta$ being a constant \cite{Blaschke:2020vuy}. The value of this parameter depend on a particular choice of the quark matter EoS. 
For the CSS parameterization holds $\delta=0.1$ \cite{Blaschke:2020vuy}. 
We independently find $\rm M_{SP}$ and $\delta$ by fitting Eq. (\ref{XXXIX}) to $\rm M_{max}$ and $\rm M_{onset}$ extracted from the mass-radius diagrams shown on Fig. \ref{fig9}. 
Note that for the $\eta_V=0$ case we consider only the mass-radius curves supporting stable quark branches. 
In this case $\rm M_{SP}=1.31 M_\odot$ and $\delta=0.06$. 
For the case of finite vector coupling shown on the lower panel 
of Fig. \ref{fig9}, we obtain $\rm M_{SP}=1.97 M_\odot$ and $\delta=0.24$. 
As is seen from Fig. \ref{fig9}, $\rm M_{SP}$ perfectly fits the special point mass at vanishing and finite $\eta_V$.

\begin{figure}[t]
\includegraphics[width=0.9\columnwidth]{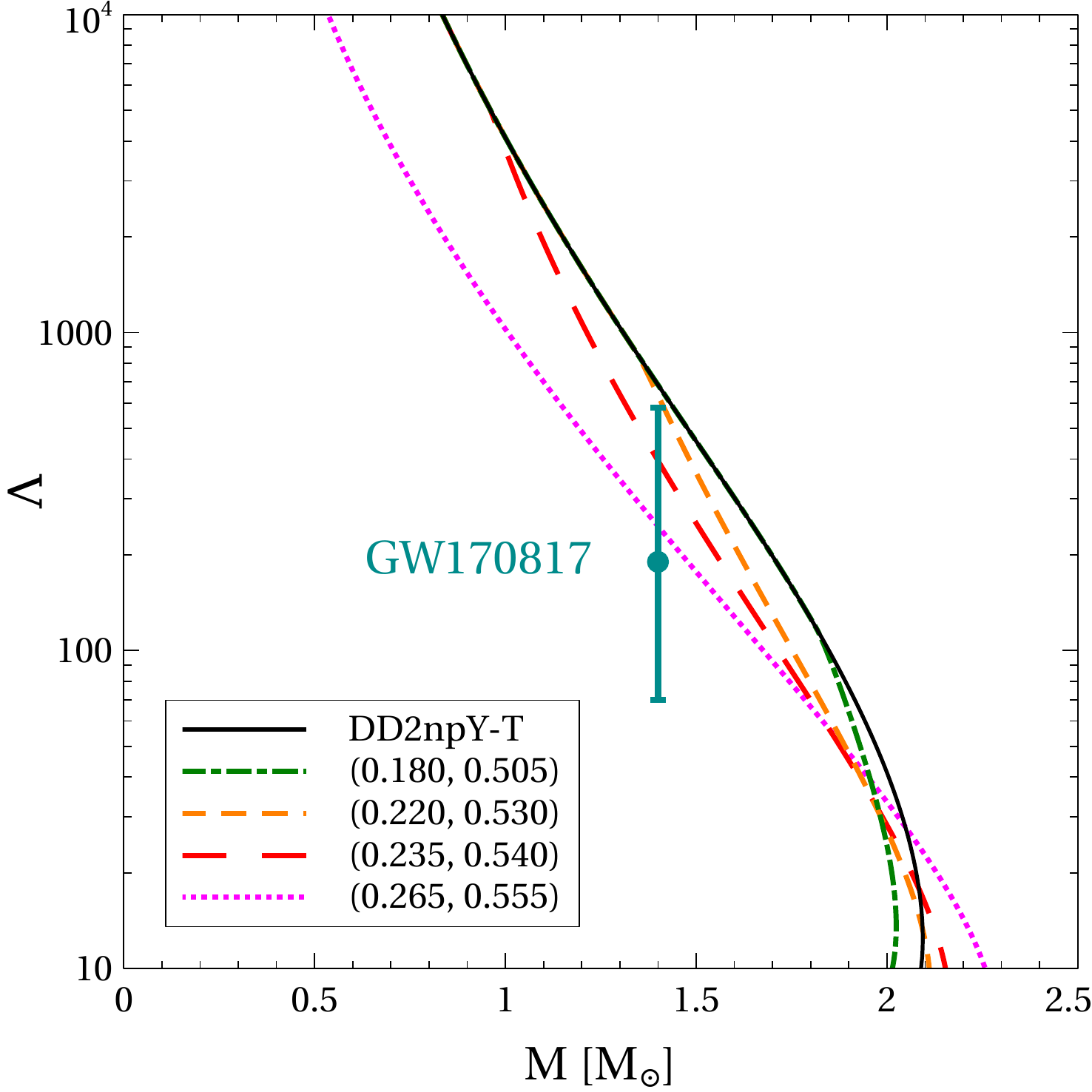}\\
\includegraphics[width=0.9\columnwidth]{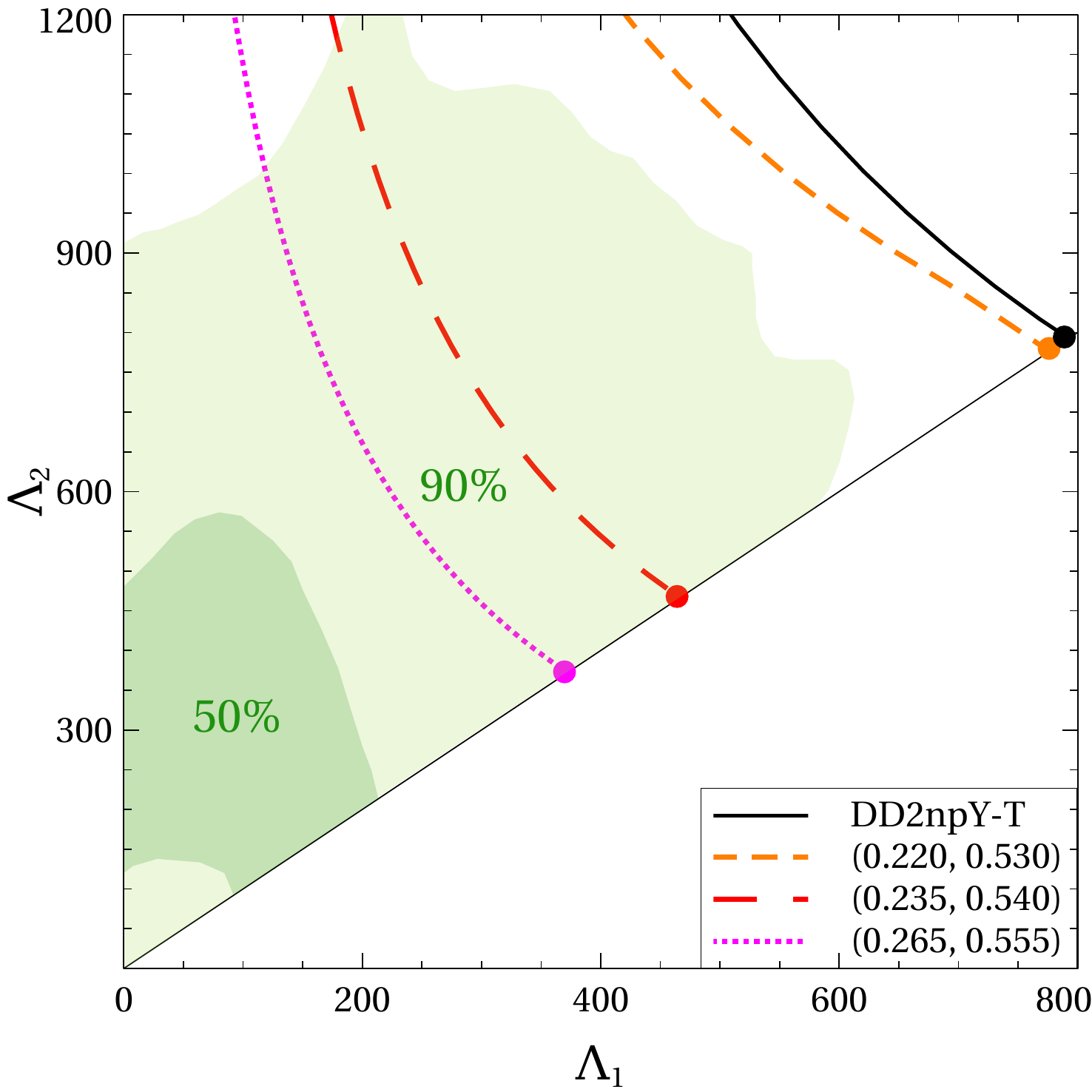}
\caption{Dimensionless tidal deformability $\Lambda$ as a function of stellar mass $\rm M$ (upper panel) and the same quantity of the low mass component of the NS-NS merger $\Lambda_2$ as a function of the corresponding parameter $\Lambda_1$ of the high mass one (lower panel). The calculations are performed for the quark-hadron EoSs presented on the lower panel of Fig. \ref{fig8}. The $\Lambda_1-\Lambda_2$ curve $(0.180,0.505)$ is not shown since within the shown region it coincides with the purely hadronic one. Colored filled circles on the lower panel represent the configurations with equal masses of two components $\rm M_1=M_2=1.3646~ M_\odot$. Dark and light green shaded areas represent the regions falling into the $50~\%$ and $90~\%$ confidence levels.}
\label{fig10}
\end{figure}

The gravitational wave signal frrom the inspiral phase of the binary neutron star merger GW170817 is another source of information about the EoS for neutron star matter. 
The most important quantity extracted form the observational data is the dimensionless tidal deformability $\Lambda=\frac{2}{3}k_2C^2$ expressed through the second Love number $k_2$ and the stellar compactness $C=\frac{\rm M}{\rm R}$. For a compact star with the mass $\rm  M=1.4~M_\odot$ it is constrained to $\Lambda_{1.4}=190^{+390}_{-120}$ \cite{LIGOScientific:2018cki}. 
At vanishing vector coupling, $\eta_V=0$, the EoS of quark matter does not support this value of the stellar mass. 
Therefore, we analyse $\Lambda$ only at finite vector coupling. The dependence of the dimensionless tidal deformability on the stellar mass is shown on the top panel of Fig. \ref{fig10}. 
The observational data constraint on $\Lambda$ is fulfilled only for sufficiently high diquark couplings providing a low onset mass of quark matter and thus a sufficient compactness of the stellar configurations. 
This constraint is fulfilled by the curves (0.235, 0.540) and (0.265, 0.555) yielding $\Lambda_{1.4}=395$ and $\Lambda_{1.4}=325$, respectively. 
The best agreement is achieved for the same model, which is the most accurate with respect to the discussed observational constraint on the mass-radius diagram.

The masses of two components $\rm M_1$ and $\rm M_2$ of the event GW170817 are connected by the chirp mass \cite{Peters:1963ux}
\begin{eqnarray}
\label{XXXX}
\mathcal{M}=\frac{(\rm M_1 M_2)^{\frac{3}{5}}}{(\rm M_1+M_2)^{\frac{1}{5}}},
\end{eqnarray}
which is an important characteristics of the merger. 
By convention $\rm M_1\ge M_2$. 
The best measured combination of the masses from GW170817 
yields $\mathcal{M}=1.188^{+0.004}_{-0.002}~M_\odot$ \cite{LIGOScientific:2017vwq}, which is used below. 
We also rely on the results of the recent analysis imposing a common EoS of two components \cite{LIGOScientific:2018cki}. Compared to the case of independently imposed EoS \cite{LIGOScientific:2018mvr}, it gives a more stringent constraint on the tidal deformabilities of two components $\Lambda_1$ and $\Lambda_2$. 
For each set of $\rm M_1$ and $\rm M_2$ providing the chosen value of $\mathcal{M}$, the tidal deformabities of two components were calculated through EoS-insensitive relations \cite{Yagi:2013awa}.
The results are demonstrated in the lower panel of Fig. \ref{fig10}. 
Increasing the diquark coupling lowers the quark matter onset mass, leading in turn to more compact stellar configurations and smaller tidal deformabilities. 
However, an increase of the vector coupling stiffens the quark matter EoS and causes the tidal deformability of the star to grow. 
Two curves fulfilling the constraint on $\Lambda_{1.4}$, 
namely the ones for the parameter sets (0.235, 0.540) and (0.265, 0.555), are well inside the 90\% confidence level and lie not so far from the 50\% one. 

\begin{figure}[t]
\includegraphics[width=0.9\columnwidth]{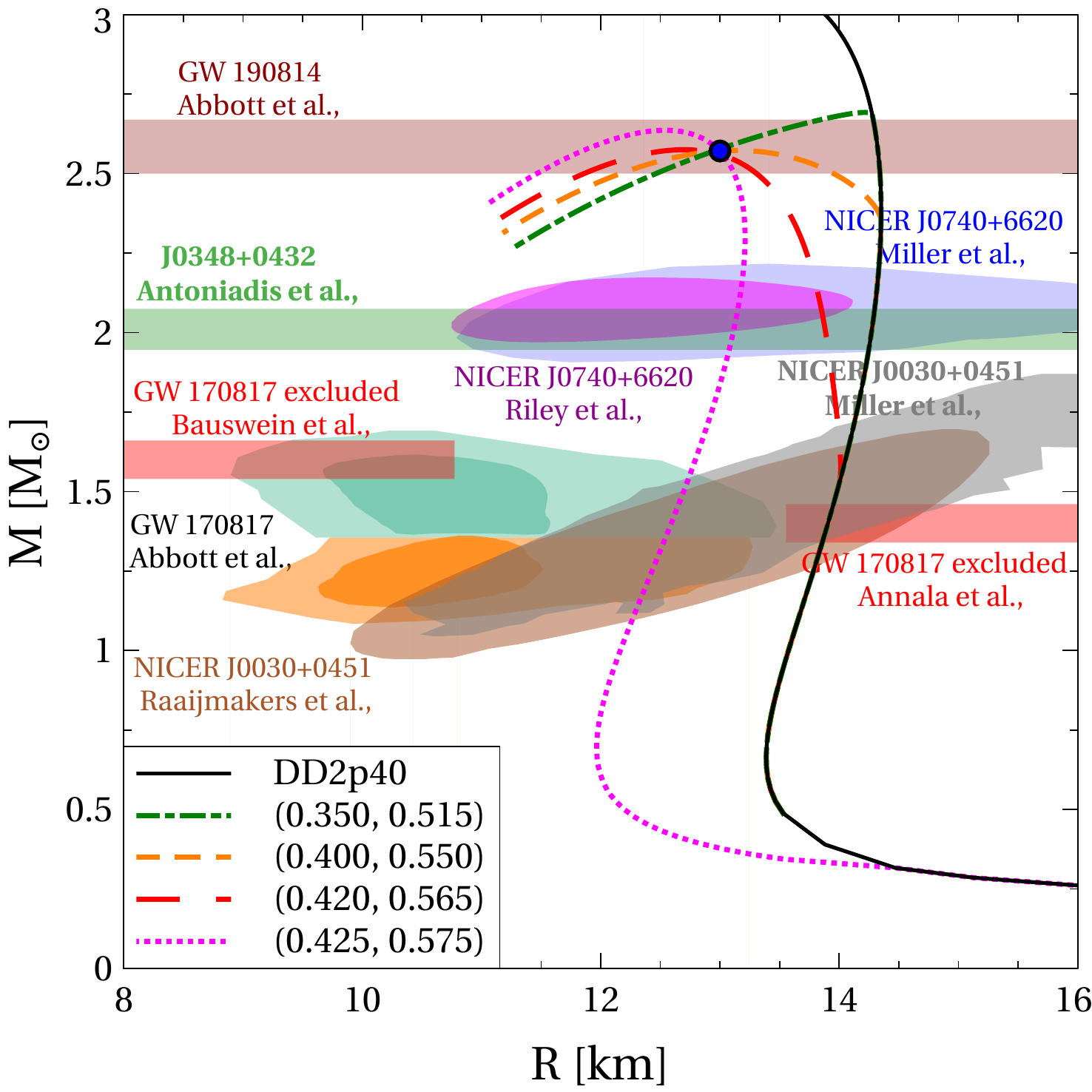}
\caption{
The same as in Fig. \ref{fig9} but with the DD2p40 model for hadron part of hybrid EoS and with different values of $\eta_V$ and $\eta_D$ given in the legend. The upper colored band represents the range of stellar masses reported in Ref. \cite{LIGOScientific:2020zkf}.}
\label{fig11}
\end{figure}

The 2.5 - 2.67 $\rm M_\odot$ compact object from the merger GW190814  is either the lightest black hole or the heaviest neutron star ever observed \cite{LIGOScientific:2020zkf}. Despite the fact that both of these possibilities are neither confirmed nor falsified, the latter one raises the question about stability of ultra heavy neutron stars. As was mentioned before, sufficiently high values of the vector and diquark couplings stabilize quark matter in the interiors of heavy compact stars. 
However, the values of $\eta_V$ and $\eta_D$ needed to provide stability of the 2.6 $M_\odot$ stars within our approach prevent the Maxwell construction of the quark-hadron transition due to absence of crossing of the quark matter EoS with the DD2npY - T hadronic one in the $\mu_B-p$ plane. It is worth mentioning, that this issue itself might be related to the phase transition construction, i.e. the Glendenning \cite{Glendenning:1992vb}, pasta \cite{Heiselberg:1992dx,Yasutake:2014oxa,Maslov:2018ghi} or interpolation \cite{Masuda:2012ed,Alvarez-Castillo:2013spa,Baym:2017whm,Blaschke:2018pva,Abgaryan:2018gqp,Ayriyan:2021prr,Ivanytskyi:2022wln} constructions. 
The corresponding analysis lays out of the scope of the present work. Therefore, we use as a stiffer hadronic EoS the DD2p40 one being a version of the DD2 EoS with the excluded volume correction \cite{Typel:2016srf} that allows the Maxwell construction at high values of $\eta_V$ and $\eta_D$. 
The mass-radius relation obtained with the corresponding hybrid EoS is shown on Fig. \ref{fig11}. Remarkably, a proper choice of the vector and diquark couplings allows our model not only to reach 2.6 $M_\odot$ stellar mass, but also makes it consistent with the stellar radius and tidal deformability constraints from the GW170817 merger \cite{LIGOScientific:2018cki}. For example, at $\eta_V=0.425$ and $\eta_D=0.575$ our approach yields $\rm M_{max} = 2.64~M_\odot$, $\rm R_{1.4}=12.6$ km and $\Lambda_{1.4}=543$. It is also worth mentioning, that these values of the vector and diquark couplings lead to a rather early onset of quark matter. The corresponding values of the baryon density are $n_B=0.17~fm^{-3}$ on the hadron boundary of the mixed phase and $n_B=0.27~fm^{-3}$ on the quark one. At the considered values of the coupling constants and for densities typical for stellar interiors, the squared speed of sound of quark matter varies within the interval $c_S^2=0.63 - 0.8$.

As is seen from Fig. \ref{fig11}, the set of hybrid EoS selected has a special point with the mass $\rm M_{SP}=2.56~M_\odot$ perfectly fitting Eq. (\ref{XXXIX}) with $\delta=0.035$. This value of $\delta$ differs from those found for the special points shown on Fig. \ref{fig9}. This supports our conclusion about the model dependence of the scaling (\ref{XXXIX}), which itself is universal. 
%

\section{Conclusions}
\label{sec8}

We have studied a relativistic density functional approach to quark matter, which 
i) mimics quark confinement by a rapid growth of the quark self-energy in the confining region, 
ii) respects chiral symmetry of strong interaction Lagrangian 
model and 
iii) can be interpreted as a chiral quark model with density dependent coupling constants. 
In addition to the vector repulsion channel, we have introduced the diquark pairing that was not studied before within the relativistic density functional approaches to quark matter.

In order to model NS with quark cores we apply the present density functional approach to Maxwell-construct a hybrid EoS made of a hadronic one and the new color superconducting quark matter EoS with the vector repulsion. The results of this modeling are confronted with the recent constraints on the NS mass-radius relation and tidal deformability from multi-messenger astronomy. 
As a general feature, we want to stress that color superconductivity lowers the onset density of quark matter and provides the present approach with the possibility to describe hybrid star sequences with a maximum mass above 2.6 $\rm M_\odot$, while being
sufficiently compact at 1.4 $\rm M_\odot$, consistent with the mentioned constraints. 
The approach has still a potential for further improvement, e.g., by adjusting the values of the vector and diquark couplings and by modifying the hadron-to-quark matter transition construction.

\section*{Acknowledgements}
This work has been supported by the Polish National Science Centre (NCN) under grant No. 2019/33/B/ST9/03059. It is part of a project that has received funding from the European Union’s Horizon 2020 research and innovation program under grant agreement STRONG – 2020 - No 824093. 

\section*{APPENDIX A}
\label{sec-app-a}
\setcounter{section}{1}
\renewcommand{\thesection}{\Alph{section}}
\setcounter{equation}{0}
\renewcommand{\theequation}{\thesection.\arabic{equation}}

Here we consider the confinement mechanism from Ref. \cite{Fowler:1981rp} in order to demonstrate how it relates to the ones from the present work and from Ref. \cite{Kaltenborn:2017hus}. 
For this we start with the zero temperature pressure of the gas of quark quasi particles with medium dependent mass $m^*$ and chemical potentials $\mu_f$
\begin{eqnarray}
\label{A1}
p=2\sum_{f,c}\int\frac{d{\bf k}}{(2\pi)^3}
\left[\epsilon_{{\bf k}f}g_{\bf k}+
(\mu_f-\epsilon_{{\bf k}f})f_{{\bf k}f}\right]-\mathcal{U}+n_s\frac{\partial \mathcal{U}}{\partial n_s},
\nonumber\\
\end{eqnarray}
where $f_{{\bf k}f}=\theta(\mu_f-\epsilon_{{\bf k}f})$, $\mathcal{U}$ is the quasiparticle interaction potential, which is a function of their scalar density $n_s$ discussed below. 
The remaining notations coincide with the ones adopted in the main text of this work. 
We note that the last two terms in Eq. (\ref{A1}) are equivalent to $-\mathcal{U}_{MF}+\langle\overline{q}q\rangle\Sigma_{MF}$ in Eq. (\ref{VIII}). 
At the moment we do not discuss the form of $\mathcal{U}$ in order to keep the analysis general. 
Contrary to the present consideration, Ref. \cite{Fowler:1981rp} neglects the interaction potential and the zero point energy term that represents the contribution of fermionic zero mode.

The condition of the pressure stationarity $\partial p/\partial m^*=0$ is equivalent to Eq. (\ref{XXII}). The definition of the scalar density
\begin{eqnarray}
\label{A2}
n_s=2\sum_{f,c}\int\frac{d{\bf k}}{(2\pi)^3}
\left[f_{{\bf k}f}-g_{\bf k}\right]\frac{m^*}{\epsilon_{{\bf k}f}}.
\end{eqnarray}
allows us to write this condition in the form $\partial m^*/\partial n_s=\partial^2 \mathcal{U}/\partial n_s^2$. It can be integrated in order to obtain the mass gap equation
\begin{eqnarray}
\label{A3}
m^*=m+\frac{\partial \mathcal{U}}{\partial n_s},
\end{eqnarray}
where the current quark mass $m$ appears as an integration constant. Eq. (\ref{A3}) along with the definition (\ref{A2}) should be solved selfconsistently in order to find $m^*$ for a given $\mathcal{U}$. 
Once this is done, the quark number density $n$ and energy density $\varepsilon$ can be found using the thermodynamic identities discussed in Section \ref{sec2}
\begin{eqnarray}
\label{A4}
n&=&2\sum_{f,c}\int\frac{d{\bf k}}{(2\pi)^3}f_{{\bf k}f},\\
\label{A5}
\varepsilon&=&2\sum_{f,c}\int\frac{d{\bf k}}{(2\pi)^3}
\left[f_{{\bf k}f}-g_{\bf k}\right]\epsilon_{{\bf k}f}+\mathcal{U}-n_s\frac{\partial \mathcal{U}}{\partial n_s}.
\end{eqnarray}
Eq. (\ref{A5}) coincides with Eq. (1) of Ref. \cite{Fowler:1981rp} if the contributions of fermionic zero mode and interaction potential are neglected. 
Let us consider the regime of vanishing quark number density. 
In this case $m^*$ is large and the low-momentum states with $\epsilon_{{\bf k}f}\simeq m^*$ are the most important. 
This allows us to approximate
\begin{eqnarray}
\label{A6}
n_s&=&n-2\sum_{f,c}\int\frac{d{\bf k}}{(2\pi)^3}g_{\bf k},\\
\label{A7}
\varepsilon&=&m^*n_s+\mathcal{U}-n_s\frac{\partial \mathcal{U}}{\partial n_s}.
\end{eqnarray}
Adopting the assumption of Ref. \cite{Fowler:1981rp} that $\varepsilon$ at vanishing density converges to some constant value $B$, we conclude from the last expression that
\begin{eqnarray}
\label{A8}
m^*=\frac{B-\mathcal{U}}{n_s}+\frac{\partial \mathcal{U}}{\partial n_s}.
\end{eqnarray}
This result demonstrates that at vanishing $\mathcal{U}$ the low density scaling $m^*=B/n$ from Ref. \cite{Fowler:1981rp} is respected only at $n_s=n$ provided by the no-sea approximation that neglects the contribution of fermionic zero mode ($g_{\bf k}=0$). 
On the other hand, setting $\mathcal{U}=0$ at any $m^*\neq m$ does not fulfill the mass gap equation (\ref{A3}). 
If it is respected, then Eq. (\ref{A7}) turns to $\varepsilon=mn_s+\mathcal{U}$, which along with $\varepsilon=B$ makes Eqs. (\ref{A3}) and (\ref{A8}) coincide. 
For the SFM potential $\mathcal{U}_{SFM}=D_0n_s^{2/3}$ this yields $m^*_{SFM}=m+(2/3)D_0n_s^{-1/3}$, which at small densities and 
$g_{\bf k}=0$ gives $m^*_{SFM}\propto n^{-1/3}$ \cite{Horowitz:1985tx,Ropke:1986qs,Kaltenborn:2017hus}. Thus, the low density scaling $m^*\propto n^{-1}$ requires the no-sea approximation and violates the mass gap equation that would be required for the pressure stationarity.

\section*{APPENDIX B}
\label{sec-app-b}
\setcounter{section}{2}
\renewcommand{\thesection}{\Alph{section}}
\setcounter{equation}{0}
\renewcommand{\theequation}{\thesection.\arabic{equation}}

Here we analyze the high density asymptotic of the 2SC phase at vanishing temperature. 
For definiteness we consider $\mu_f^*>0$. 
In this regime $\mu_f^*\gg\Lambda\gg m_f^*$ and quark masses can be neglected as well as zero point terms. 
In addition, only quarks ($a=+$) with 
$f_{{\bf k}fc}^+=\theta(\mu_f^*-|{\bf k}|)$ 
contribute to the thermodynamic quantities. 
Then, the pairing gap equation becomes
\begin{eqnarray}
\label{B1}
\Delta\simeq-4G_D\sum_{f,c}\int\frac{d{\bf k}}{(2\pi)^3}
f_{{\bf k}fc}^+\frac{\Delta_c}{\epsilon_{{\bf k}fc}^+}.\,
\end{eqnarray}
At $\Delta\sim\mu_f^*$ or at $\Delta\ll\mu_f^*$, the right hand side of this equation diverges as $\sim\Delta\sum_f\mu_f^{*2}$, which is much faster than $\Delta$ corresponding to the left hand side. 
Therefore, $\Delta\gg\mu_f^*$. 
In this case the upper edge of the filled single particle states lies below the Fermi surface, 
$\epsilon_{{\bf k}fc}^+\simeq-\Delta$, and Eq. (\ref{B1}) 
becomes
\begin{eqnarray}
\label{B2}
\Delta\simeq4G_D\sum_{f,c}\int\frac{d{\bf k}}{(2\pi)^3}
f_{{\bf k}fc}^+\frac{\Delta_c}{\Delta}=\frac{4G_D}{3\pi^2}\sum_f\mu_f^{*3}.
\end{eqnarray}
Similarly, the equation for the vector field simplifies to 
\begin{eqnarray}
\label{B3}
\omega\simeq-4G_V\sum_{f,c}\int\frac{d{\bf k}}{(2\pi)^3}f_{{\bf k}fc}^+
\left(2\Delta_c\delta(\epsilon_{{\bf k}fb}^+)+\frac{\epsilon_{{\bf k}fb}^+}{\epsilon_{{\bf k}fc}^+}\right).
\end{eqnarray}
For paired quarks, the bracket in this expression can be approximated by 
$2\Delta\delta(\epsilon_{{\bf k}fb}^+)$ since $\frac{\epsilon_{{\bf k}fb}^+}{\epsilon_{{\bf k}fc}^+}\ll 1$, while $f_{{\bf k}fc}^+$ can be replaced by one half due to the Dirac delta function. 
The corresponding momentum integral behaves as $\sim\Delta\sum_f\mu_f^{*2}$. 
The contribution of unpaired blue quarks can be neglected since it behaves as $\sim\sum_f\mu_f^{*3}$. 
Thus, with $\epsilon_{{\bf k}fb}^+\simeq|{\bf k}|-\mu_f^*$ the vector field becomes 
\begin{eqnarray}
\label{B4}
\omega\simeq-4G_V\Delta \sum_{f,c=r,g}\int\frac{d{\bf k}}{(2\pi)^3}\delta(\epsilon_{{\bf k}fb}^+)=-\frac{4G_V\Delta}{\pi^2}\sum_f\mu_f^{*2}.
\end{eqnarray}
The quark number density is
\begin{eqnarray}
\label{B5}
\langle q^+q\rangle=-\frac{\omega}{2G_V}\simeq\frac{2\Delta}{\pi^2}\sum_f\mu_f^{*2}.
\end{eqnarray}
This gives a direct access to number density of a given quark flavor $\langle f^+ f\rangle\simeq2\Delta\mu_f^{*2}/\pi^2$. 

In order to find the asymptotic value of the effective quark mass we also consider the chiral condensate. 
Applying the simplifications adopted before and $\epsilon_{{\bf k}f}\simeq|{\bf k}|$ one gets
\begin{eqnarray}
\label{B6}
\langle\overline{q}q\rangle\simeq4\Delta m^*\sum_f\int\frac{d{\bf k}}{(2\pi)^3}
\frac{\delta(\epsilon_{{\bf k}fb}^+)}{|{\bf k}|}=
\frac{2\Delta m^*}{\pi^2}\sum_f\mu_f^*.
\end{eqnarray}
Using definition of the effective quark mass, we obtain
\begin{eqnarray}
\label{B7}
m^*=m-2G_{PS}\langle\overline{q}q\rangle\simeq
\frac{m}{1+\frac{4G_{PS}\Delta}{\pi^2}\sum\limits_f\mu_f^*}.
\end{eqnarray}

The quark pressure can be found using Eq. (\ref{XVII}) as $p_q=-\Omega_q+\Omega_{q0}$, 
with $\Omega_{q0}$ being quark thermodynamic potential in the vacuum. 
For this we integrate by parts, neglect $\Omega_{q0}$, zero point and antiquark terms and arrive to 
\begin{eqnarray}
\label{B8}
p_q\simeq2\sum_{f,c}\int\frac{d{\bf k}}{(2\pi)^3}
f^+_{{\bf k}fc}\left(2\Delta_c\delta(\epsilon_{{\bf k}fb}^+)+\frac{\epsilon_{{\bf k}fb}^+}{\epsilon_{{\bf k}fc}^+}\right)
\frac{{\bf k}^2}{3\epsilon_{{\bf k}f}}.
\end{eqnarray}
The contribution of paired quarks can be found similar to the case of $\omega$. It is $\frac{2\Delta}{3\pi^2}\sum_f\mu_f^{*3}\simeq
\frac{\Delta^2}{2G_D}$, where Eq. (\ref{B2}) was used. 
The blue unpaired quarks contribute to $p_q$ as $\sim\frac{1}{12\pi^2}\sum_f\mu_f^{*4}$, which can be neglected. Thus, the total pressure becomes 
\begin{eqnarray}
\label{B9}
p\simeq p_q+\frac{\omega^2}{4G_V}-\frac{\Delta^2}{4G_D}\simeq
\frac{\omega^2}{4G_V}+\frac{\Delta^2}{4G_D}.
\end{eqnarray}
The energy density is expressed using the corresponding thermodynamic identity and the definition of the effective chemical potential
\begin{eqnarray}
\label{B10}
\varepsilon=\sum_f\langle f^+ f\rangle\mu_f-p\simeq
\frac{\omega^2}{4G_V}+\frac{5\Delta^2}{4G_D}.
\end{eqnarray}
At finite vector coupling $\omega\gg\Delta$ and $\varepsilon\simeq p\simeq\frac{\omega^2}{4G_V}$, while at $G_V=0$, one gets $\omega=0$ and $\varepsilon\simeq 5p\simeq\frac{5\Delta^2}{4G_D}$. 
This yields the asymptotic value of the squared speed of sound
\begin{eqnarray}
\label{B11}
c_S^2=\frac{dp}{d\varepsilon}\simeq\left\{
\begin{array}{l}
1,\quad G_V\neq0\\
\frac{1}{5},\quad G_V=0
\end{array}\right..
\end{eqnarray}

Now we focus on the electrically neutral case of $\beta$-equilibrium at $G_V\neq0$. Using above mentioned expressions for number density of $u$ and $d$ quarks along with the electron number density $n_e=\mu_e^3/(3\pi^2)$ the corresponding condition becomes
\begin{eqnarray}
\label{B12}
\frac{2}{3}\frac{2\Delta\mu_u^{*2}}{\pi^2}-
\frac{1}{3}\frac{2\Delta(\mu_u^*+\mu_e)^2}{\pi^2}-
\frac{\mu_e^3}{3\pi^2}=0.
\end{eqnarray}
where $\mu_d^*=\mu_u^*+\mu_e$ is used in order to express effective chemical potential of $d$ quarks. The electron term in Eq. (\ref{B12}) is negligibly small compared to the quark ones since $\mu_e\sim\mu_f^*\ll\Delta$. With this we find $\mu_e\simeq\left(2^{1/2}-1\right)\mu_u^*$ and $\mu_d^*\simeq2^{1/2}\mu_u^*$. Inserting this $\mu_d^*$ to Eqs. (\ref{B2}) and (\ref{B4}) we find vector field $\omega\sim\mu_u^{*5}$ and effective chemical potential of $u$ quarks
\begin{eqnarray}
\label{B13}
\mu_u^*\simeq\mu_u-\frac{16G_DG_V}{3\pi^4}\mu_u^{*5}.
\end{eqnarray}
Two terms on the right hand side of this equation diverge faster then $\mu_u^*$ on the left hand side, which can be neglected. This yields $\mu_u^*\sim\mu_u^{1/5}$. We also approximate chemical potential of $u$ quarks as $\mu_u=\mu_B/3-2\mu_e/3\simeq\mu_B/3$, which is valid since $\mu_e\sim{\mu_u}^{1/5}\ll\mu_u$. Thus
\begin{eqnarray}
\label{B14}
\mu_u^*\simeq
\left(\frac{\pi^4 \mu_B}{16G_DG_V}\right)^{\frac{1}{5}},
\end{eqnarray}
With this equation and $\mu_d^*=2^{1/2}\mu_u^*$ asymptotic expressions for pairing gap and quark number density become
\begin{eqnarray}
\label{B15}
\Delta&\simeq&\frac{4G_D}{3\pi^2}\left(1+2^{\frac{3}{2}}\right)
\left(\frac{\pi^4 \mu_B}{16G_DG_V}\right)^{\frac{3}{5}},\\
\label{B16}
\langle q^+q\rangle&\simeq&
\frac{1+2^{\frac{3}{2}}}{2G_V}\mu_B.
\end{eqnarray}
%

\bibliography{RDFM}

\end{document}